\documentclass{ptephy_v1}

\preprintnumber{XXXX-XXXX}

\usepackage{boites}
\usepackage{graphics}

\begin{document}

\title{Acceleration and propagation of ultra high energy cosmic rays}


\author{\name{\fname{Roberto} \surname{Aloisio}}{1,2,*}}
\address{
\affil{1}{$^1$Gran Sasso Science Institute, viale F. Crispi 7, 67100 L'Aquila, Italy\\}
\affil{2}{$^2$INFN - Laboratori Nazionali del Gran Sasso, via G. Acitelli 22, Assergi, L'Aquila, Italy}
\email{roberto.aloisio@gssi.it}
}

\begin{abstract}
We review the physics of the highest energy cosmic rays. The discovery of their sources, still unknown, will reveal the most energetic astrophysical objects in the universe and could unveil new physics beyond the standard model of particle physics. We discuss the details of propagation of these high energy particles, their interaction with astrophysical photon backgrounds and intergalactic magnetic fields, and the production of secondary cosmogenic particles associated to their transport. We examine different models of acceleration, reviewing the principal astrophysical objects that could energise cosmic rays until the highest energies. Given the uncertainties in the observed mass composition, we review the two alternative scenarios of a composition made only by protons or by protons and heavier nuclei; discussing the consequences of the two scenarios in terms of sources, acceleration mechanisms and production of secondary cosmogenic gamma rays and neutrinos. 
\end{abstract}

\subjectindex{high energy astrophysics, cosmic ray physics}

\maketitle

\tableofcontents

\newpage

\section{Introduction}
\label{sec:intro}

Ultra High Energy Cosmic Rays (UHECR), namely Cosmic Rays with energies larger than $10^{17}\div 10^{18}$ eV, are the most energetic particles observed in nature, with energies exceeding $10^{20}$ eV. The first observation of particles with such extreme energies dates back to the Volcano Ranch experiment in 1962 \cite{Linsley:1963km}. Since then, the study of UHECR became one of the most important research fields in high energy Astrophysics and Astroparticle Physics, connected with the most energetic phenomena in the universe it could unveil new physics in regimes not testable otherwise. Starting from the sixties several experiments were operated with the aim of increasing the statistics of the observed UHECR events, in order to perform high resolution measurements of the flux, mass composition and anisotropy (for a review discussing the first generation detectors see \cite{Nagano:2000ve}). 

Nowadays, the most advanced experiments to detect UHECR are the Pierre Auger Observatory in Argentina \cite{ThePierreAuger:2015rma}, far the largest experimental setup devoted to the study of UHECR, and the Telescope Array (TA) experiment \cite{AbuZayyad:2012kk,Tokuno:2012mi,Tinyakov:2014lla}, placed in the United States, with roughly $1/10$ of the Auger statistics. 

Both detectors exploit the hybrid concept, combining an array of surface detectors (SD) to sample Extensive Air Showers (EAS, shower) when they reach the ground and telescopes, overlooking the surface array, to collect the fluorescence light of the atmospheric nitrogen excited (fluorescence detectors, FD). The advent of the hybrid approach has been a major breakthrough in the detection of UHECR since this method allows to have the same energy scale in the surface detectors and the fluorescence telescopes. In the past, the absence of an energy scale common to both detection methods had led to the puzzle about the existence of the flux suppression around $5\times 10^{19}$ eV, which was observed by HiRes~\cite{Abbasi:2007sv}, through FD, but not present in AGASA data~\cite{Takeda:2002at}, whose energy calibration was based on SD and Monte Carlo simulations. The first hybrid measurements were done in HiRes/MIA~\cite{AbuZayyad:1999xa} with a detector array of limited size. The Auger project, for the first time, adopted the hybrid approach~\cite{Abreu:2010aa} as the basis of the detector design solving the suppression puzzle.

The experimental study of UHECR clarified few important characteristics of these particles: (i) UHECR are charged particles with a limit on photon and neutrino fluxes around $10^{19}$ eV at the level of few percent and well below respectively \cite{Abraham:2009qb,Abu-Zayyad:2013dii,Abreu:2013zbq}, (ii) the spectrum observed at the Earth shows a slight flattening at energies around $5\times 10^{18}$ eV (called the ankle) with (iii) a steep suppression at the highest energies \cite{ThePierreAuger:2013eja,Abu-Zayyad:2013qwa}.

The composition of UHECR is still matter of debate. Before the advent of Auger the experimental evidences were all pointing toward a light composition with a proton dominated flux until the highest energies, sources injecting soft spectra and acceleration energies larger than $10^{20}$ eV \cite{Berezinsky:2002nc}. The measurements carried out by the Auger observatory \cite{Aab:2014kda} have shown that the mass composition of CRs, from prevalently light at $\sim 10^{18}$ eV, becomes increasingly heavier towards higher energies. Several independent calculations \cite{Allard:2011aa,Fang:2012rx,Aloisio:2013hya,Aab:2016zth} showed that spectrum and composition observed by Auger can be well explained only if sources  provide very hard spectra and a maximum rigidity $\sim 5\times 10^{18}$ V. If confirmed, these findings would represent a change of paradigm respect to the picture of ten years ago. On the other hand, the TA experiment, even if with $1/10$ of the Auger statistics, collected data that seem to confirm the pre-Auger scenario \cite{Abbasi:2014sfa}, the mass composition is compatible with being light for energies above $10^{18}$ eV, with no apparent transition to a heavier mass composition. 

As we discuss in section \ref{sec:models}, a joint working group made of members of both collaborations, TA and Auger, has recently concluded that the results of the two experiments are not in conflict once systematic and statistical uncertainties have been taken into account \cite{Abbasi:2015xga}. This conclusion, though encouraging on one hand, casts serious doubts on the possibility of reliably measuring the mass composition at the highest energies, unless some substantially new piece of information becomes available. In section \ref{sec:conclu} we will come back to the observation of mass composition, discussing the future prospects of these measurements. It should be also noted that the spectra measured by the two experiments, though being in general agreement, differ beyond the systematic error at the highest energies (where mass composition differs the most) in such a way that TA claims a spectral suppression at $\gtrsim 5\times 10^{19}$ eV while Auger shows the suppression at sensibly lower energies \cite{Verzi:2015dna}. 

In order to interpret the observations at the Earth, trying to constrain the possible sources of UHECR, it is very important a detailed modelling of UHECR propagation in the intergalactic medium, which is mainly conditioned by the interaction with astrophysical backgrounds and intergalactic magnetic fields. These interactions, that we discuss in section \ref{sec:prop}, shape the spectrum observed at the Earth and are also responsible for the production of secondary (cosmogenic) particles: photons and neutrinos. This secondary radiation, discussed in section \ref{sec:models}, can be observed through ground-based or satellite experiments and brings important informations about the mass composition of UHECR and, possibly, on their sources. 

Sources of UHECR are still a mystery, we do not know which kind of astrophysical object is responsible for the production of these particles. As we discuss in section \ref{sec:source}, there are basically two different classes of mechanisms that, in principle, could be invoked to accelerate UHECR. The first class is based on the transfer of energy from a macroscopic object (that can move relativistically or not) to microscopic particles through repeated interactions with magnetic inhomogeneities. The second class is based on the interaction with electric fields that, through high voltage drops, can accelerate particles until the highest energies. 

The transfer of energy from a macroscopic object to a particle was elaborated already by Fermi in 1949 \cite{Fermi:1949ee}, the so-called Fermi mechanism. The interaction of particles with magnetic scattering centres of a macroscopic moving object led to an energy gain per interaction of the order of $\Delta E/E \propto \beta^2$, being $\beta$ the average velocity of the scattering centres in units of $c$. This process is called second order Fermi mechanism. The first order mechanism is realised in the case of a coherent macroscopic motion, as for shock waves where particles gain energy bouncing back and forth through the shock front with an energy gain in each passage $\Delta E/E\propto \beta$ \cite{1977ICRC11.132A,Bell:1978fj,Blandford:1978ky,Blasi:2014roa}, being $\beta$ the shock's velocity in ubits of $c$. These kind of mechanisms all require many interactions to achieve the extreme energies observed. 

Shock waves are frequent in the universe, typically produced when ejected material from an astrophysical object encounters, with supersonic motion, the surrounding medium with the principal effect of heating the medium. Depending on the velocity of the ejected material, shocks can be either non-relativistic or relativistic and both kinds of shocks can be used to construct acceleration models for UHECR. To this purpose, as we discuss in section \ref{sec:source}, interesting shock regions are Gamma Ray Bursts (GRB), jets and hot-spots in Active Galactic Nuclei (AGN) or large scale accretion shocks in galaxy clusters. 

Acceleration of charged particles through electric fields is in principle very efficient. However, plasmas, ubiquitous in any astrophysical environment, typically destroy large scale electric fields in the universe reducing the effectiveness of such mechanisms. As we discuss in section \ref{sec:source}, only in particular environments high voltage drops can survive, such as in the case of magnetosphere or winds in neutron stars or near black holes and their accretion disks. 

As we discuss in section \ref{sec:exotic}, the production of UHECR could also be connected with new physics not testable in Earth's laboratories. This is the case of models in which UHECR are directly produced at high energy, as decay products of super-heavy relic particles (with mass $M_X \gtrsim 10^{13}\div 10^{14}$ GeV) predicted in a wide class of inflationary scenarios. These models connect the observations of UHECR with the Dark Matter (DM) problem and the physics of the early universe. 

The theoretical interest in UHECR is also connected with their kinematic regime. The interactions of these particles are characterised by huge values of the Lorentz factor (up to $\Gamma\simeq 10^{11}$ for protons), not reached in any other (known) physical environment. As we discuss in section \ref{sec:exotic}, the kinetic regime of UHECR provides unprecedented tests of the relativity principle and the fundamental ideas beyond the theory of special relativity. 

The paper is organised as follows. In section \ref{sec:prop} we discuss the propagation of UHECR through astrophysical backgrounds and intergalactic magnetic fields. In section \ref{sec:models} we discuss the comparison of theoretical expectations with the observations of Auger and TA, illustrating the constraints that experimental data already imply on the characteristics of sources. In section \ref{sec:source} we discuss different classes of astrophysical sources and the acceleration mechanisms related to them. In section \ref{sec:exotic} we discuss the connection between the physics of UHECR and exotic models beyond the standard model of particle physics. Finally, in section \ref{sec:conclu} we conclude discussing and summarising the principal results. 

\section{Transport of ultra-high energy cosmic rays}
\label{sec:prop}

The extra-galactic origin of UHECR, at least at energies above the ankle $E>10^{19}$ eV, is widely accepted \cite{Aloisio:2007rc,Aloisio:2012ba}. The propagation of UHECR through the intergalactic medium is conditioned primarily by astrophysical photon backgrounds and, if any, by the presence of magnetic fields. The astrophysical backgrounds involved are the Cosmic Microwave Background (CMB) and the Extra-galactic Background Light (EBL). 

The CMB, relic radiation from the big bang, is the background with the highest density and a well known cosmological evolution, while EBL, composed of infrared, optical and ultraviolet photons produced and reprocessed by astrophysical sources, has a less understood cosmological evolution, typically model dependent. In the past years several models for the cosmological evolution of EBL have been proposed \cite{Franceschini:2008tp,Stecker:2006eh,Stecker:2005qs,Kneiske:2003tx}. These models show sizeable differences only at high redshift ($z > 4$), not actually relevant in the propagation of UHECR but affecting the production of secondary cosmogenic particles \cite{Allard:2011aa,Aloisio:2015ega}, as we discuss in section \ref{sec:sec}.

\subsection{Interactions of ultra-high energy cosmic rays}
\label{sec:protons-int}

The interactions of UHECR (protons\footnote{Here we do not consider the case of neutrons because their decay time is much shorter than all other scales involved in the propagation of UHECR \cite{Aloisio:2008pp,Aloisio:2010he}.} or heavier nuclei) with astrophysical backgrounds give rise to the processes of: pair-production, photo-pion production and, only in the case of nuclei heavier than protons, photo-disintegration. Moreover, protons propagation is affected only by the CMB while for nuclei, and only in the case of photo-disintegration, also the EBL plays a role \cite{Aloisio:2008pp,Aloisio:2010he}. 

Soon after the CMB discovery, these mechanisms of energy losses were discussed \cite{Penzias:1965wn}. Most important was the so called Greisen, Zatsepin and Kuzmin (GZK) feature\footnote{Also referred as the GZK cut-off or suppression.}: a sharp suppression in the spectrum of UHECR protons expected at $E_{GZK}\simeq 5\times 10^{19}$ eV, due to the photo-pion production process on the CMB. At the highest energies also the flux of nuclei is suppressed by the photo-disintegration process on the CMB, producing the Gerasimova and Rozental (GR) cut-off \cite{GR} expected at energies $E_{GR}\simeq A\Gamma_c m_N$ being: $\Gamma_c=3\div 5 \times 10^{9}$ a critical Lorentz factor weakly dependent on the nucleus type, $A$ is the atomic mass number of the nucleus and $m_N$ the proton mass. Other important studies were conducted by Hillas \cite{Hillas:1967} and Blumenthal \cite{Blumenthal:1970nn}, on the effects of the pair production process, and by Berezinsky and Zatsepin \cite{Beresinsky:1969qj,Stecker:1973sy,Strong:1973,Berezinsky:1975zz}, on the production of secondary cosmogenic gamma rays and neutrinos. 

The rate of interactions (probability per unit time) suffered by UHECR can be written in a very general form as \cite{Stecker:1968uc,Aloisio:2008pp}:
\begin{equation}
\frac{1}{\tau} = \frac{c}{2\Gamma^2}\int_{\epsilon'_{\min}}^{+\infty} \epsilon'\sigma(\epsilon')\int_{{\epsilon'}/{2\Gamma}}^{+\infty} \frac{n_\gamma(\epsilon)}{\epsilon^2} d\epsilon \,d\epsilon' \label{eq:tau}~,
\end{equation} 
where $\Gamma$ is the Lorentz factor of the particle, $\sigma(\epsilon')$ is the total cross-section associated to the interaction, $\epsilon'$ is the background photon energy in the particle rest frame, $\epsilon'_{\min}$ is the lowest value of~$\epsilon'$ above which the interaction is kinematically possible (threshold), and $n_\gamma(\epsilon)\,d\epsilon$ is the number per unit volume of background photons with energy between $\epsilon$ and $\epsilon + d\epsilon$ in the laboratory reference frame. The photon energy in the particle rest frame is related to that in the laboratory frame by 
$$\epsilon' = \Gamma\epsilon(1-\cos\theta) \qquad \qquad 0 \le \epsilon' \le 2\Gamma\epsilon ~,$$ 
where $\theta$ is the angle between the particle and photon momenta.

The physical processes responsible for energy losses depend on the particle type and energy. Protons with Lorentz factor 
$$\Gamma\gtrsim \frac{m_{\pi}}{\epsilon(1-\cos\theta)} \qquad \qquad \frac{m_\pi}{2\epsilon} \simeq 10^{10} \lesssim \Gamma < \infty$$ 
interacting with CMB photons give rise to the photo-pion production process: 
\begin{equation}
p(n)+\gamma \to p(n) + \pi^0   \qquad p(n)+\gamma \to n(p) + \pi^+(\pi^-).
\end{equation}

At lower energies $\Gamma\lesssim10^{10}$ the same processes can occur on the EBL field \cite{Aloisio:2015ega}, although with a lower probability. The photo-pion production process implies a sizeable energy loss for protons resulting in the GZK suppression of the flux \cite{Greisen:1966jv,Zatsepin:1966jv} which arises at the threshold energy that, in the nucleon rest frame, reads 
$$\epsilon'_{\min} = m_{\pi} + \frac{m_{\pi}^2}{2m_N} \approx 145 ~{\rm MeV.}$$

The cross-section of protons' photo-pion production has a complex behaviour with a number of peaks corresponding to different hadronic resonances, the largest one being the $\Delta$ resonance placed at $\epsilon' =\epsilon_{\Delta}\approx 340$ MeV \cite{Berezinsky:2002nc}. At energies much larger than $\epsilon_{\Delta}$ the cross-section has an approximately constant value \cite{Berezinsky:2002nc}. 

\begin{figure}[!h]
\centering
\includegraphics[scale=.45]{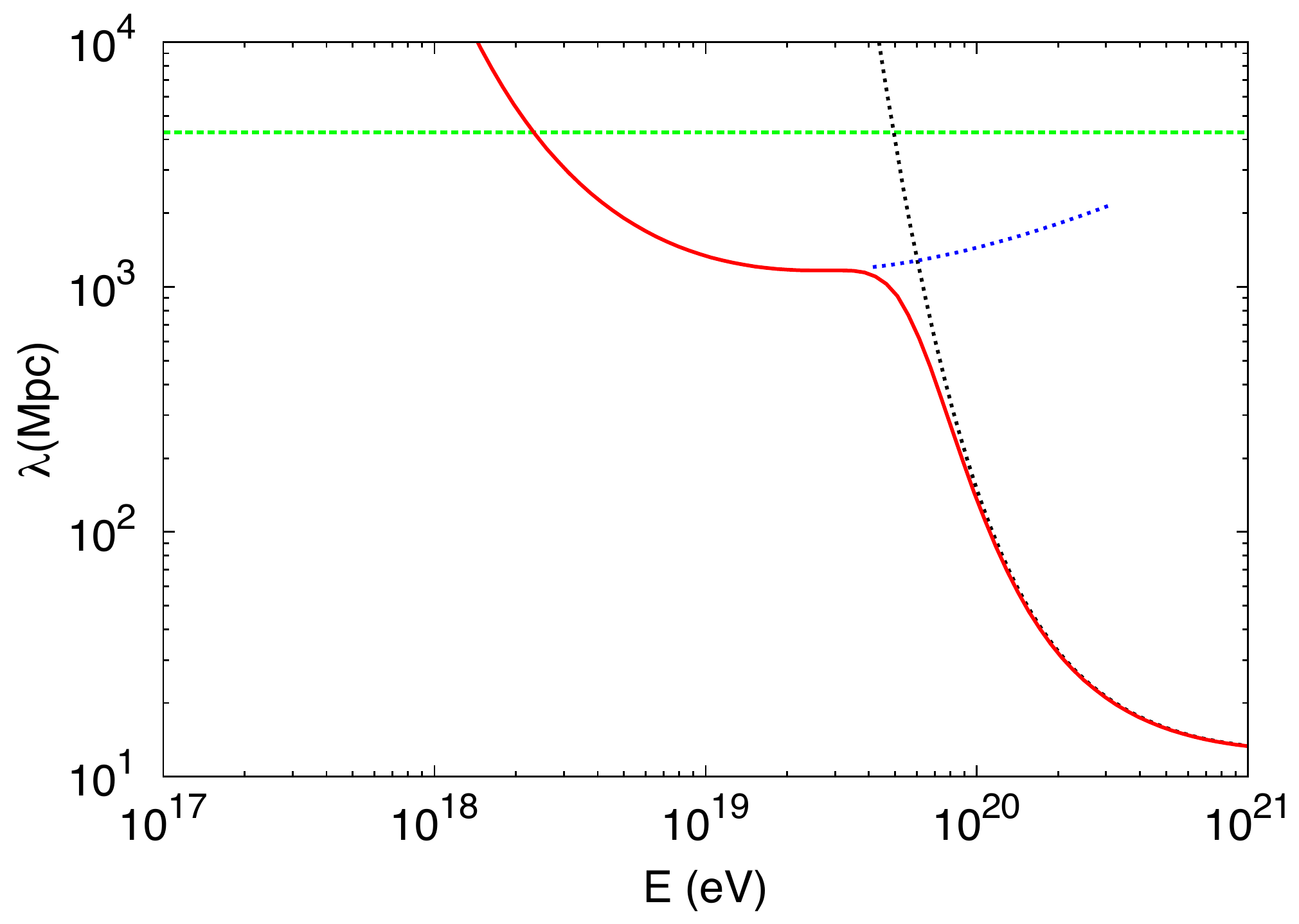}
\caption{Loss length of UHE protons (red solid line total energy losses). Photo-pion production dotted black line, pair production dotted blue line. The size of the visible universe is seen by the dashed green line.}
\label{fig1}   
\end{figure}

The photo-pion production process holds also for nucleons bound within UHE nuclei, being the interacting nucleon ejected from the parent nucleus. This process is subdominant if compared with photo-disintegration except at extremely high energies \cite{Allard:2011aa}. As we discuss in section \ref{sec:models}, it has some relevance only in the case of the production of secondary cosmogenic particles. 

UHE nuclei interacting with CMB and EBL photons can be stripped of one or more nucleons (photo-disintegration process):
\begin{equation}
(A,Z) + \gamma \to (A-n, Z-n') + nN
\end{equation}
being $A$ and $Z$ the atomic mass number and atomic number of the nucleus, $n$ ($n'$) the number of stripped nucleons ($N$). In the nucleus rest frame the energy involved in such processes is usually much less than the rest mass of the nucleus itself, therefore in the laboratory frame all fragments keep approximately the same Lorentz factor of the parent nucleus, i.e. we can neglect nucleus recoil \cite{Aloisio:2008pp,Aloisio:2010he}. 

The cross-section is dominated by a smooth peak, the giant dipole resonance, that appears for photon energies close to the threshold ($8~{\rm MeV} \approx \epsilon'_{\min} < \epsilon' \le 30~{\rm MeV}$) \cite{Puget:1976nz}, in this regime photons interact with "all" nucleons that behave collectively as a fluid. The giant dipole resonance corresponds to the extraction of one nucleon and it is the dominating process in UHE nuclei propagation \cite{Puget:1976nz,Allard:2005ha,Aloisio:2008pp,Aloisio:2010he}. At larger energies $\epsilon'>30~{\rm MeV}$ the quasi-deuteron process dominates, in which the photon interacts with one or two nucleons inside the nucleus with the extraction of two or more nucleons. This regime corresponds to an almost constant cross-section and has a small impact on the propagation of UHE nuclei \cite{Puget:1976nz,Allard:2005ha,Aloisio:2008pp,Aloisio:2010he}. 

The process of photo-disintegration is responsible for the production of secondary hadronic particles, stripped away from the primary injected nuclei. Therefore, injecting heavy nuclei at the source produces a flux at the Earth composed also by lighter nuclei emitted during propagation. 

\begin{figure}[!h]
\centering\includegraphics[scale=0.58]{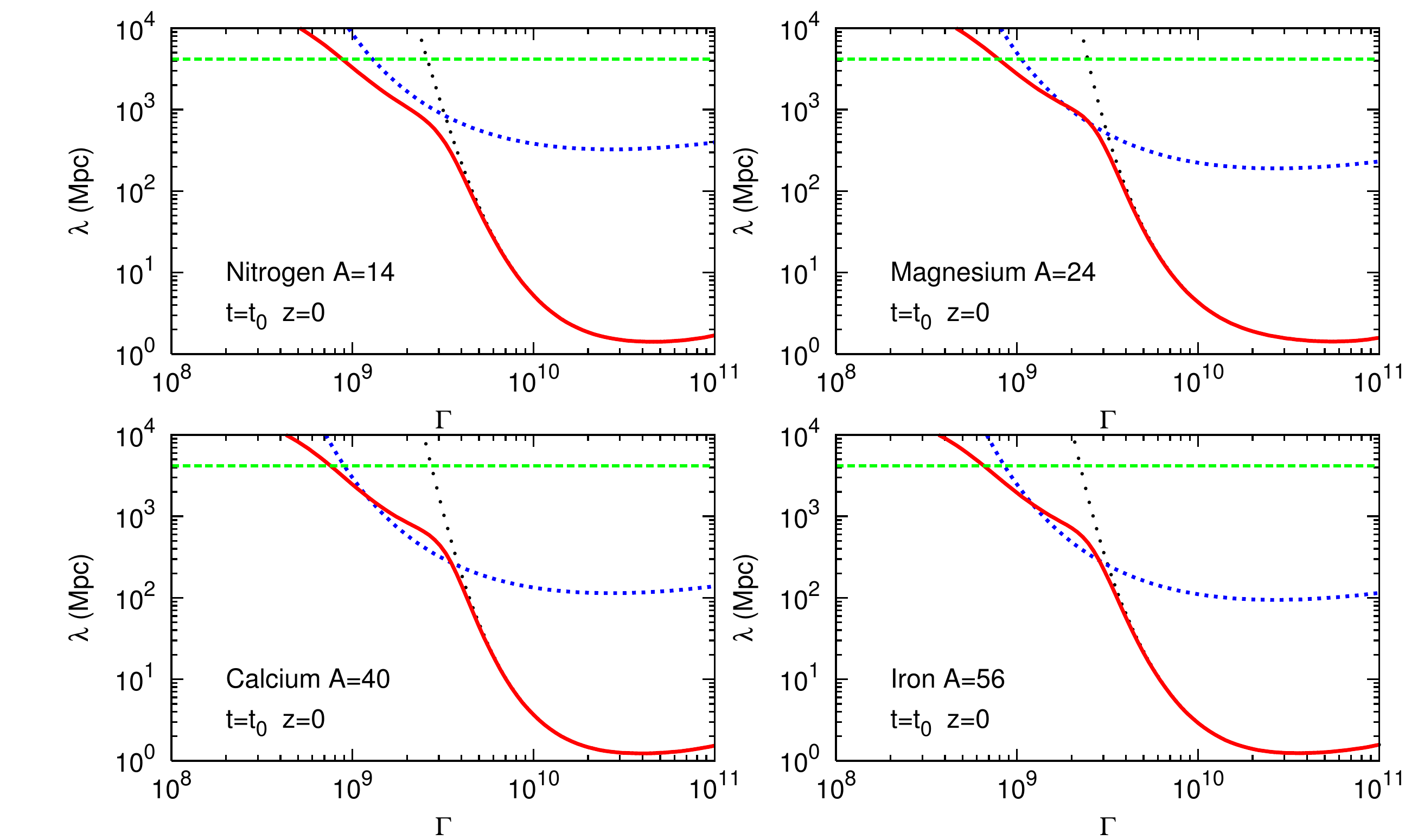}
\caption{Loss length of four species of UHE nuclei (as labeled). Red solid line photo-disintegration and blue dotted line pair-production. The effect of EBL (modelled as in \cite{Stecker:2005qs}) on photo-disintegration is seen by the black dotted line that shows photo-disintegration due to the sole CMB field. The size of the visible universe is seen by the dashed green line (adiabatic energy losses). Figures taken from \cite{Aloisio:2010he}.}
\label{fig2}   
\end{figure}

Protons and nuclei with Lorentz factor $\Gamma\ge 10^{9}$ can undergo the process of pair production: 

$$p(A) + \gamma \to p(A) + e^{+} + e^{-}$$ 

The mean free path associated to this interaction is relatively short compared with all other length scales of propagation, with a very small amount of energy lost by the propagating particle in each interaction \cite{Berezinsky:2002nc}. Taking into account this result, we can always interpret pair production as a process that continuously depletes the particle's energy. Hence, the rate of energy losses $\frac{1}{\Gamma}\left (\frac{d\Gamma}{dt}\right)$ due to pair-production can be written substituting in Eq. (\ref{eq:tau}) $\sigma \to \sigma f$, being $f$ the inelasticity of the process, i.e. the average fraction of energy lost by the particle in one interaction \cite{Ginzburg:1990sk}. In the case of nuclei, the rate of pair-production energy losses can be computed starting from the rate of protons and using the scalings \cite{Aloisio:2008pp,Aloisio:2010he}:  

$$ f^A=f^p/A \qquad {\rm and} \qquad \sigma^A=Z^2\sigma^p ~.$$ 

Particles covering cosmological distances feel the effect of the changes in the background universe due to cosmology. Thus suffering adiabatic energy losses. Assuming standard cosmology we can write the energy lost per unit time by UHECR (protons or nuclei) as 
\begin{equation}
\left(-\frac{1}{\Gamma}\frac{d\Gamma}{dt}\right)_\textrm{ad} = H(z) = H_0\sqrt{(1+z)^3\Omega_\textrm{m} + \Omega_\Lambda}
\label{adiabatic}
\end{equation}
where $z$ is the redshift at time~$t$, $H_0 \simeq 70~\mathrm{km}/\mathrm{s}/\mathrm{Mpc}$ is the Hubble constant, $\Omega_\textrm{m} \simeq 0.26$ is the matter density, and $\Omega_\Lambda \simeq 0.74$ is the dark energy density \cite{Ade:2015xua}.

In figures \ref{fig1}, \ref{fig2} we plot the interaction path length of protons and nuclei computed at zero red-shift as function of energy and Lorentz factor respectively.

In the case of protons (figure \ref{fig1}), at low energies, up to few $10^{18}$ eV, the energy losses are dominated by the expansion of the universe. The pair production process starts to be relevant at the threshold energy $2\times 10^{18}$ eV; the pion photo-production becomes important at $\simeq 5\times 10^{19}$ eV (GZK cut-off) where the loss length drops to very low values and the large scale universe becomes opaque to UHECR protons. 

In figure \ref{fig2}, the loss length of different nuclei is plotted as function of the Lorentz factor $\Gamma$, assuming the EBL model presented in \cite{Stecker:2005qs}. Two sharp drops can be seen in the path length of nuclei; the first drop, at energy $A\times 10^{18}$ eV ($\Gamma\simeq 10^{9}$), is the combination of photo-disintegration on far infra-red photons (low energy EBL) and pair-production on CMB; the second drop, even more pronounced, is due to photo-disintegration on CMB photons and arises at energies $E_{GR} \simeq A\times 4\times 10^{18}$ eV (GR cut-off). 

\subsection{Propagated spectra}
\label{sec:prospec}
The spectrum of CR observed at the Earth always consists in the combination of what was injected by sources and the effects of propagation. Once the interactions of UHECR are specified one can determine the spectra expected at the Earth by assuming an injection spectrum at the sources and their cosmological evolution\footnote{We will not discuss here the case of extra-galactic magnetic fields, we will come back to this point in the forthcoming section \ref{sec:mag}.}. In the following we will assume that sources have a uniform and homogeneous distribution, with emissivity\footnote{Energy emitted (in UHECR) per unit time and volume. Given a distribution of sources with number density $n_S$ each with the same luminosity $L_S$ the energy emitted per unit time and volume is given by ${\cal L}_S=n_SL_S$.} ${\cal L}_S$ eventually dependent on red-shift, all share the same injection spectral index $\gamma_g$ (with a power law injection $\propto E^{-\gamma_g}$) and the same maximum energy, with a rigidity dependent behaviour $E_{max}(A,Z)=Z\times E_{max}^p$. Under these simplified assumptions, the propagation of UHECR can be described by the set of differential equations \cite{Aloisio:2008pp,Aloisio:2010he}: 

\begin{equation}
\frac{\partial n_p(\Gamma,t)}{\partial t} - \frac{\partial}{\partial
\Gamma} \left [ b_p(\Gamma,t)n_p(\Gamma,t) \right ] =  Q_p(\Gamma,t)
\label{eq:kin_p}
\end{equation}   
\begin{equation}
\frac{\partial n_{A}(\Gamma,t)}{\partial t} - \frac{\partial}{\partial
\Gamma} \left [ n_{A}(\Gamma,t) b_{A}(\Gamma,t) \right ] +
\frac{n_{A}(\Gamma,t)} {\tau_{A}(\Gamma,t)}  = Q_A (\Gamma,t)
\label{eq:kin_A}
\end{equation}
where $n$ is the equilibrium distribution of particles, $b=-d\Gamma/dt$ is the rate of decrease of the particle's Lorentz factor, $Q_p$ and $Q_A$ are the production rates per unit co-moving volume and time of protons and nuclei, as the sum of those produced by the sources and the secondary products of photo-disintegration. Equations (\ref{eq:kin_p}) and (\ref{eq:kin_A}) are valid under the hypothesis of continuum energy losses\footnote{As discussed for the pair production process, in the hypothesis of continuum energy losses the rate $1/\Gamma (d\Gamma/dt)$ can be determined substituting $\sigma\to f\sigma$ in Eq. (\ref{eq:tau}) being $f$ the inelasticity of the process (see above) as presented in \cite{Ginzburg:1990sk}.}, i.e. we neglect the effects of stochasticity in the interactions, and modelling photo-disintegration as a "decaying" process (with "life time" $\tau_A$) that just depletes the number of photo-disintegrated nuclei \cite{Aloisio:2008pp,Aloisio:2010he}. 

As discussed above, the leading channel of photo-disintegration is the extraction of one nucleon $(A+1)+\gamma \to A + N$. The primary nucleus, the secondary nucleus and the associated (emitted) nucleon all have approximately the same Lorentz factor $\Gamma$, therefore the injection rate of secondary particles can be written as:
\begin{equation} 
Q^{\rm sec}_A(\Gamma,t)=Q^{\rm
sec}_p(\Gamma,t)=\frac{n_{A+1}(\Gamma,t)}{\tau_{A+1}(\Gamma,t)}.
\label{eq:Q_sec}
\end{equation} 

Equation (\ref{eq:Q_sec}) couples together the transport equations for nuclei, Eqs. (\ref{eq:kin_A}), that should be solved following the photo-disintegration chain, namely starting from the solution for the primary injected nucleus with mass number $A_0$  (injected at the source with $Q_{A_0}(\Gamma,t)$) and then using the solution to solve the equation for the nucleus with mass number $A_0-1$ (injection $Q^{sec}_{A_0-1}=n_{A_0}/\tau_{A_0}$), moving downward along the photo-disintegration chain until the lowest mass secondary nucleus $A=2$. The equilibrium distribution of protons $n_p$ is obtained by solving Eq. (\ref{eq:kin_p}) with both the injection of freshly accelerated protons and secondary protons produced by photo-disintegration:

$$Q_p(\Gamma,t)=Q_p^{\rm acc}(\Gamma,t)+\sum_{A<A_0}Q_{p}^{\rm sec}(\Gamma,t)~.$$

We complete the discussion above by presenting the analytical solution to equations (\ref{eq:kin_p}) and (\ref{eq:kin_A}), written in integral form, as function of redshift $z$ and particles' Lorentz factor $\Gamma$. Following \cite{Aloisio:2010he,Aloisio:2008pp} the solution reads:
\begin{equation}
J_p(\Gamma,z)=\frac{c}{4\pi}\int_{z}^{z_{\rm max}}  \frac{dz'}{(1+z)H(z)}
Q_p^(\Gamma',z') \left (\frac{d\Gamma'}{d\Gamma}\right )_p,
\label{eq:np}
\end{equation}

\begin{equation}
J_{A}(\Gamma,z)=\frac{c}{4\pi}\int_{z}^{z_{max}} \frac{dz'}{(1+z') H(z')}
Q_{A}(\Gamma ',z')\left ( \frac{d\Gamma '}{d\Gamma} \right )_A
e^{-\eta_{A}(\Gamma ',z')},
\label{eq:nA0-solut}
\end{equation}

where $\Gamma'(z')$ is the particles' Lorentz factor at redshift $z'$ (with $\Gamma'=\Gamma$ at $z'=z$), $d\Gamma'/d\Gamma$ for protons and nuclei is calculated in \cite{Berezinsky:1988wi,Aloisio:2010he,Aloisio:2008pp}. The term $e^{-\eta_A}$ in equation (\ref{eq:nA0-solut}) takes into account the effect of photo-disintegration of the propagating nucleus $A$, with the photo-disintegration treated as a "decay" process it reads:
\begin{equation} 
\eta_A(\Gamma',z') = \int_z^{z'} \frac{dz''}{(1+z'') H(z'')}
\frac{1}{\tau_A(\Gamma'',z'')} ~,
\label{eq:etaA}
\end{equation}
being $\tau_A$ the photo-disintegration "life-time" computed as in Eq. (\ref{eq:tau}).

\subsection{Intergalactic magnetic fields} 
\label{sec:mag}

The propagation of UHECR can be largely affected by the presence of intergalactic magnetic fields\footnote{We will not discuss galactic magnetic fields that do not affect spectrum and mass composition of UHECR but can only change the arrival directions of particles \cite{Stanev:1996qj,Harari:2000az,Prouza:2003yf,Yoshiguchi:2003mc,Kachelriess:2005qm,Takami:2005ij}.} (IMF). As a general remark it should be stated that our knowledge of these fields is still poor and fragmented, even if several important constraining observations were achieved (see \cite{Kronberg:1993vk,Grasso:2000wj,Carilli:2001hj,Kulsrud:2007an,Beck:2011he,2012SSRv..166....1R,Durrer:2013pga} and references therein). In certain environments of the universe such as galaxy clusters, which could harbour sources of UHECR, the magnetic field is better known with typical values in the range of $1 ~\mu$G \cite{Kronberg:1993vk,Grasso:2000wj,Carilli:2001hj,Kulsrud:2007an,Beck:2011he,2012SSRv..166....1R,Durrer:2013pga}. 

It is outside clusters, far the largest space traversed by extra-galactic CR, in filaments and voids, that the value of the magnetic field is not known and, to date, no convincing mechanism to produce strong fields over very large (supra-cluster) scales has been clearly found. The most reliable observations of IMF are those of synchrotron emission, its polarisation and Faraday rotation at radio frequencies ($0.1\div 10$ GHz) \cite{Kronberg:1993vk,Grasso:2000wj,Carilli:2001hj,Kulsrud:2007an,Beck:2011he,2012SSRv..166....1R,Durrer:2013pga}. These measurements imply an upper limit for the IMF that depends on the assumed coherence length of the field itself. For instance, according to \cite{Blasi:1999hu}, in the case of an inhomogeneous universe $B < 4$ nG with a scale of coherence of about $l_c = 50$ Mpc. 

Apart from observations, the IMF can be predicted, in principle, implementing Magneto-Hydrodynamics (MHD) evolution of magnetic fields in numerical simulation of Large Scale Structure (LSS) formation \cite{2002A&A...387..383D,Dolag:2004kp,Sigl:2004yk,Donnert:2008sn,Ryu:2008hi}. The main ambiguities in these simulations are related to the assumed seed magnetic fields, to the mechanism invoked in their growth and to the capability of reproducing the local density velocity field (constrained \cite{2002A&A...387..383D,Dolag:2004kp} and unconstrained simulations \cite{Sigl:2004yk}). Unfortunately, because of these uncertainties, MHD simulations are not completely conclusive. The volume filling factor of strong fields of the order of $1$ nG vary by several orders of magnitude from one simulation another. The predicted magnetic field in voids (filaments) vary from $10^{-3}$ nG ($10\div 1$ nG) \cite{Dolag:2004kp,Donnert:2008sn,Ryu:2008hi} up to $10^{-1}$ nG ($10$ nG) \cite{Sigl:2004yk}.

In order to discuss the effects of magnetic fields on the propagation of UHECR let us consider the ideal configuration of a homogeneous turbulent magnetic field with strength $B$ on the coherence scale $l_c$. A charged particle with energy $E$ in a magnetic field $B$ has a Larmor radius given by:
\begin{equation}
r_L(E) \simeq 1 \left (\frac{E_{EeV}}{ZB_{nG}}\right ) Mpc
\label{eq:rL}
\end{equation}
being $E_{EeV}$ the energy in units of $10^{18}$ eV, $B_{nG}$ the magnetic field in units of nano-Gauss and $Z$ the charge of the particle. 

Depending on their energy, particles can feel the effect of the magnetic field in different ways: the scale that governs such interaction is the field coherence length $l_c$. If $r_L<l_c$, thus for energies 
$$E<E_c\simeq 10^{18} Z B_{nG} \left (\frac{l_c}{Mpc} \right )~eV,$$ 
particles, resolving the turbulence spectrum of the magnetic field, will diffuse through resonant scattering on the field components at the scale $r_L$. This regime is called quasi-linear or resonant diffusion. The diffusion length $l_D$, i.e. the distance that corresponds to a typical particle deflection of $1$ rad, depends on the type of turbulence met by the particle at the Larmor radius scale, hence it depends on the turbulence spectrum of the magnetic field. It can be easily shown that $l_D\simeq l_c (E/E_c)^{\alpha}$ \cite{Aloisio:2004jda,Lemoine:2004uw,Aloisio:2008tx,Mollerach:2013dza} being $\alpha$ related to the spectrum of magnetic turbulence ($\alpha=1/3$ Kolmogorov, $\alpha=1/2$ Kraichnan and $\alpha=1$, i.e. $l_D=r_L$, in the case of Bohm spectrum). Particles propagation in the quasi-linear regime is purely diffusive with a diffusion coefficient 

$$D=\frac{1}{3} cl_D \simeq \frac{1}{3} c l_c\left (\frac{E}{E_c} \right )^{\alpha}. $$  

At high energy as soon as $r_L>l_c$  ($E>E_c$) the scattering is non-resonant and particles propagation results in a series of small deflections $\delta\theta\simeq l_c/r_L$ in each coherence length $l_c$. The diffusion length $l_D$ can be easily evaluated as the space traversed by a particle to suffer a deflection $\Delta\theta\simeq 1$ rad, one has $l_D\simeq l_c (E/E_c)^2$ with a diffusion length that strongly increases with energy \cite{Aloisio:2004jda,Aloisio:2008tx,Mollerach:2013dza}. In this case the propagation of particles can be either diffusive or rectilinear depending on the sources distribution, namely on the average distance between sources $r_s$. If $l_D<r_s$ the propagation is still diffusive, with a diffusion coefficient 

$$D=\frac{1}{3} cl_D \simeq \frac{1}{3} c l_c \left (\frac{E}{E_c} \right )^2,$$ 

while at the highest energies when $l_D>r_s$ the rectilinear propagation regime is recovered and the magnetic field has no effects on propagation \cite{Aloisio:2004jda,Aloisio:2008tx}. 

It is important to mention here, as a general remark, that the effect of the magnetic field on spectrum and mass composition of UHECR is closely tied to the assumptions made about density and luminosity of sources. In other words, even a very strong field would have no effect on any observable if the mean separation between sources is smaller than all propagation length scales, i.e. diffusion and energy losses lengths \cite{Aloisio:2004jda,Aloisio:2008tx}.   

Typical density expected for UHECR sources is in the range $n_s=10^{-4}\div 10^{-6}$ Mpc$^{-3}$, with a rather large uncertainty it can be estimated by the observations (or non-observation) of small-scale clustering in the arrival directions \cite{Dubovsky:2000gv,Fodor:2000yi,Blasi:2003vx,Kachelriess:2004pc}. The typical separation distance between sources can be estimated as $r_s=(n_s)^{-1/3}\simeq 10\div 10^2$ Mpc.

As discussed above, assuming a non-resonant diffusion regime ($r_L > l_c$), cosmic rays from a source at distance $r_s$ will diffuse as soon as $l_D\le r_s$, or in terms of energy:
\begin{equation}
E\le E_D\simeq 10^{18} Z B_{nG}\sqrt{\frac{r_s l_c}{{\rm Mpc}^2}} eV~.
\label{eq:ED}
\end{equation}

Particles with energies below $E_D$, traveling for longer time respect to rectilinear propagation, will loose more energy. The net effect of this kind of transport is to reduce the expected flux at energies below $E_D$. 

As first realised in \cite{Aloisio:2004fz,Lemoine:2004uw}, the (IMF induced) flux suppression (in the range $10^{17}\div 10^{18}$ eV) has a paramount importance in tagging the transition between galactic and extra-galactic CR \cite{Aloisio:2012ba}. Such scenario can also be invoked to reduce the flux of protons below $10^{18}$ eV in the case of the dip model (see section \ref{sec:dip}) or to allow for softer injection spectra in the case of mixed composition (see section \ref{sec:mix}). As follows from Eq. (\ref{eq:ED}), the viability of these scenarios clearly depends on the assumptions made about the magnetic field configuration that should be at the $nG$ level. It is worth recalling here that, as follows from equipartition, a $nG$ field is an absolute upper limit of the expected magnetic field strength in voids. Considering more realistic configurations with inhomogeneous magnetic fields, i.e. taking into account different field intensities and coherence scales in voids and filaments, the suppression effect on the low energy flux will be less pronounced and shifted to lower energies \cite{Kotera:2007ca,Kotera:2008ae,Kalli:2011ud}.

Let us conclude this part by stating that the interaction of UHECR with magnetic fields can be very important also in the source environment. The strong connection between matter and magnetic fields in the universe could imply a rich phenomenology, largely oversimplified here, due to the likely presence of UHECR sources in those environments characterised by high values of the magnetic field. For instance, these effects could be at the base of source models, as we discuss in the following section, that provide a mixed mass composition. 

\section{Observations, source models and cosmogenic messengers}
\label{sec:models}

To constrain the basic characteristics of UHECR sources we adopt a purely phenomenological approach in which sources are homogeneously and isotropically distributed with the basic parameters: $\gamma_g$ injection power law index, $E_{max}$ maximum energy at the source (rigidity dependent), ${\cal L}_S$ emissivity and relative abundances of different elements at injection. These parameters are fitted to experimental data (both spectrum and mass composition) with as little as possible {\it a priori}  theoretical prejudice on what the values should be. In the forthcoming section \ref{sec:source} we will come back to the physics of sources, discussing specific astrophysical objects that show the expected characteristics. 

Solving the transport\footnote{In this section we neglect the effect of magnetic fields.} equations (\ref{eq:kin_p}) and (\ref{eq:kin_A}) we can determine the theoretical flux, to be compared with observations, while mass composition is inferred from the mean value of the depth of shower maximum $\langle X_{max} \rangle$ and its dispersion (RMS) $\sigma(X_{max})$, computed as shown in \cite{Abreu:2013env} and compared with observations too. The combined analysis of $\langle X_{max} \rangle$ and $\sigma(X_{max})$, even if not conclusive, as we discuss in section \ref{sec:conclu}, allows to obtain less model dependent information on the mass composition \cite{Aloisio:2007rc,Kampert:2012mx}. A relevant source of uncertainties comes from the dependence of the depth of shower maximum and its fluctuations on the hadronic interaction model used to describe the shower development in the atmosphere. Most of such models fit low energy accelerator data while providing somewhat different results when extrapolated to the energies of relevance for UHECR (for a review see \cite{Engel:2011zzb} and references therein).

As discussed in the Introduction, the main spectral features of UHECR observed at the Earth are: (i) the ankle, a flattening of the spectrum at energy around $\simeq 5\times 10^{18}$ eV, observed since 1960s (Volcano Ranch experiment \cite{Linsley:1963}) and confirmed by all observations \cite{Aab:2015bza,Abbasi:2015bha}, and (ii) a sharp suppression of the spectrum at the highest energies. The energy of such suppression is not univocally identified with some tension between the observations of Auger and TA: the first detector places the suppression at $E\simeq 2.5\times 10^{19}$ eV \cite{Aab:2015bza} while the second at larger energies $E\simeq 6\times 10^{19}$ eV \cite{Ivanov:2015pqx}. 

\begin{figure}[!h]
\centering\includegraphics[scale=0.27]{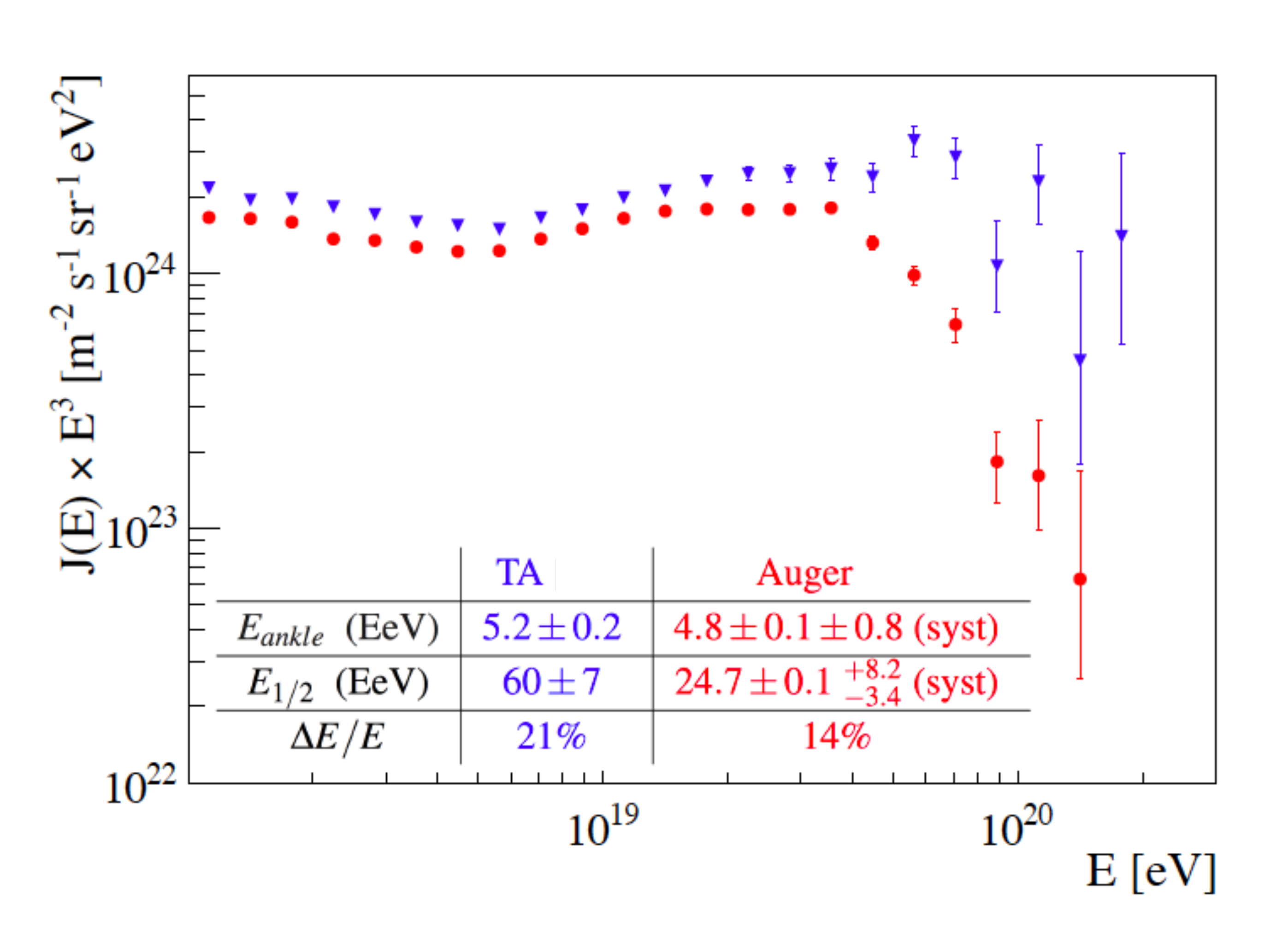}
\centering\includegraphics[scale=0.6]{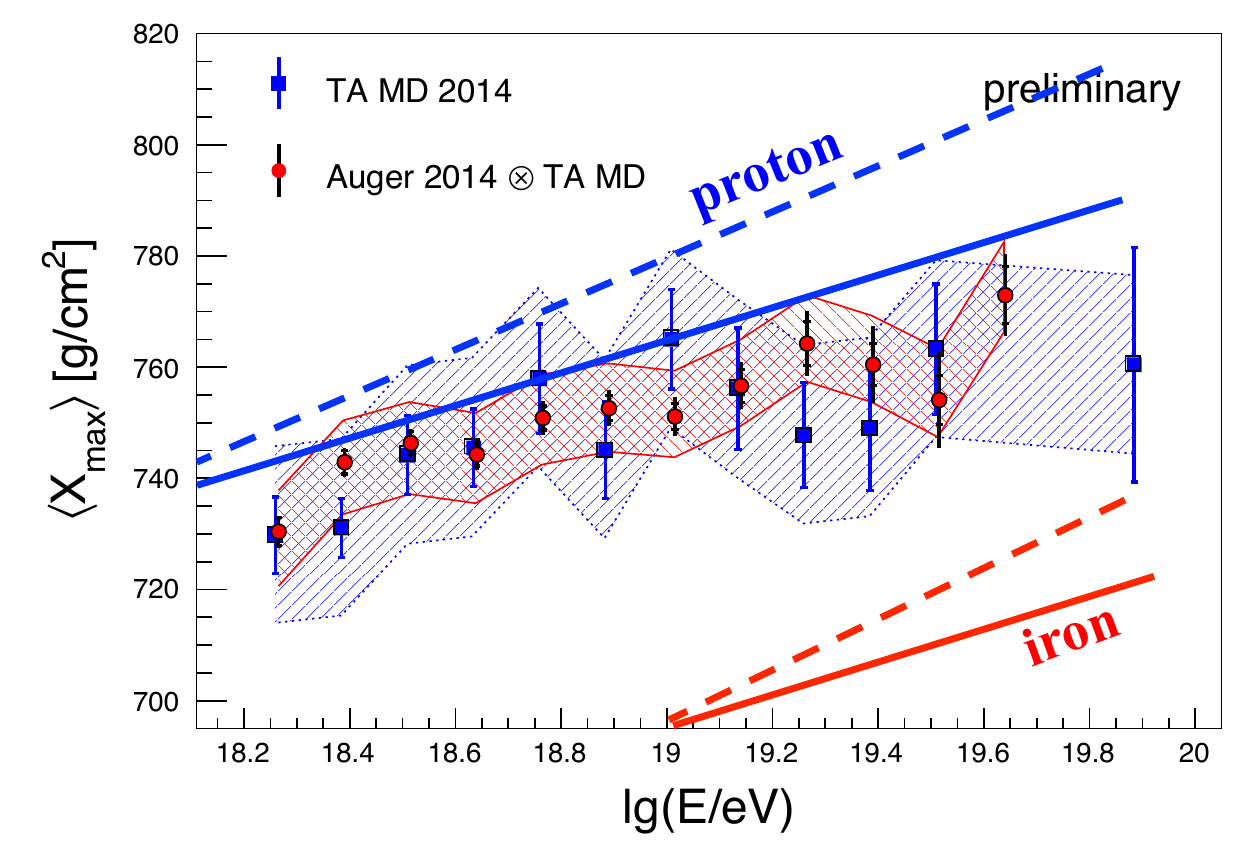}
\caption{[Left Panel] Energy spectra of UHECR as released in 2015 by Auger (red points) and TA (blue points). Also labeled, in both datasets, are: the energy of the ankle $E_{ankle}$, the energy of the high energy suppression $E_{1/2}$ and the systematics in energy determination (figure taken from \cite{Verzi:2015dna}). [Right Panel] Comparison of the $\langle X_{max} \rangle$ as obtained by the Auger-TA working group \cite{Abbasi:2015xga} with superimposed (by the author) the theoretical expectation of pure proton or iron composition (as labeled) in the case of QGS-Jet-II-03 (solid lines) and Sybil 2.1 (dashed lines) hadronic interaction models (taken from figure 30 of \cite{Abbasi:2014sfa}).}
\label{fig2bis}   
\end{figure}

In the left panel of figure \ref{fig2bis} we plot the energy spectra observed by Auger and TA as presented at the last International Cosmic Rays Conference \cite{Verzi:2015dna}, also labeled are the position of the ankle and the high energy suppression as measured in the two datasets. Apart from a shift in the energy determination, which can be reabsorbed in systematics uncertainties as labeled, it follows that the two observations seem discrepant at the highest energies in both shape and position of the suppression  \cite{Verzi:2015dna}. 

In the right panel of figure \ref{fig2bis} we plot $\langle X_{max} \rangle$ as determined by the comparison of the Auger and TA datasets by the joint working group to study mass composition \cite{Abbasi:2015xga}. Superimposed to experimental data are plotted (by the author) also the theoretical expectations for pure protons and pure irons in the two cases of QGSJet and Sybill interaction models, as extracted from figure 30 of \cite{Abbasi:2014sfa}. The result of the joint working group, as follows from figure \ref{fig2bis} (right panel), states the compatibility of the two datasets whose physical interpretation highly depends on the hadronic interaction model assumed. Hence we can conclude that the observations of Auger and TA are currently not providing an unambiguous measurement of the mass composition and, in what follows, we consider separately the two cases of a pure proton composition and a mixed composition with heavy nuclei contributing to the flux. 

\subsection{Protons and the dip model} 
\label{sec:dip}

In the case of a pure proton composition the only relevant astrophysical background is the CMB \cite{Aloisio:2008pp,Aloisio:2010he}. This fact makes the propagation of UHE protons free from the uncertainties related to the background, being the CMB exactly known as a pure black body that evolves with redshift trough its temperature. In this case any signature of the propagation in the observed spectrum can be easily referred to the assumptions made at the source, subtracting the effects of the interactions suffered during propagation. In order to isolate these effects it is useful the so-called modification factor $\eta(E)$ defined as the ratio \cite{Berezinsky:2002nc}:
  
\begin{equation}
\eta(E)=\frac{J_p(E)}{J_{unm}(E)}
\label{eq:modfact}
\end{equation}
where $J_p$ is the protons spectrum, computed with all energy losses taken into account (see Eq. (\ref{eq:np})), and $J_{unm}(E)$ is the unmodified spectrum computed taking into account only adiabatic energy losses due to the expansion of the universe. 

\begin{figure}[!h]
\centering\includegraphics[scale=.45]{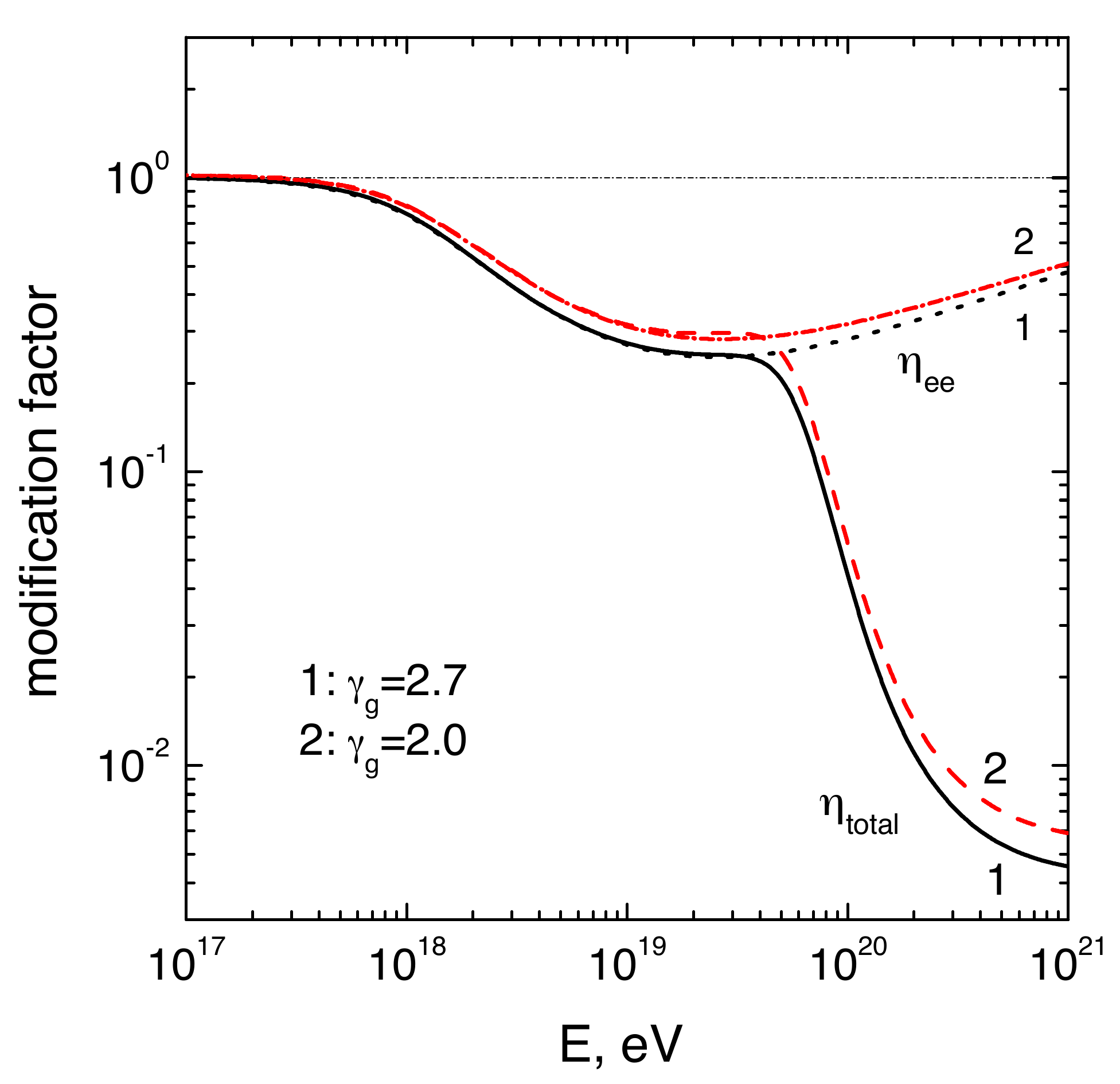}
\caption{Theoretical modification factor computed for different values of the injection power law index as labeled. Figure taken from \cite{Aloisio:2006wv}.}
\label{fig3}   
\end{figure}

In figure \ref{fig3} we plot the modification factor $\eta(E)$ as computed for different values of the injection power law index $\gamma_g$. From this figure it is evident the weak dependence of $\eta(E)$ on the injection chosen while it clearly shows the signatures of the energy losses suffered by protons. At low energy, losses are dominated by the adiabatic expansion of the universe, the spectrum at the Earth keeps the injection shape, and the modification factor is $\eta(E)=1$; above the pair production threshold (around $2\times 10^{18}$ eV) pair production energy losses become important and the propagated spectrum hardens (the curves labeled with $e^{+}e^{-}$ take into account only pair-production); above the photo-pion production threshold (around $5\times 10^{19}$ eV) the propagated spectrum experiences an abrupt steepening which corresponds to the GZK feature (curves labeled with ${\it total}$ take into account all relevant energy losses). 

\begin{figure}[!h]
\centering\includegraphics[scale=1.2]{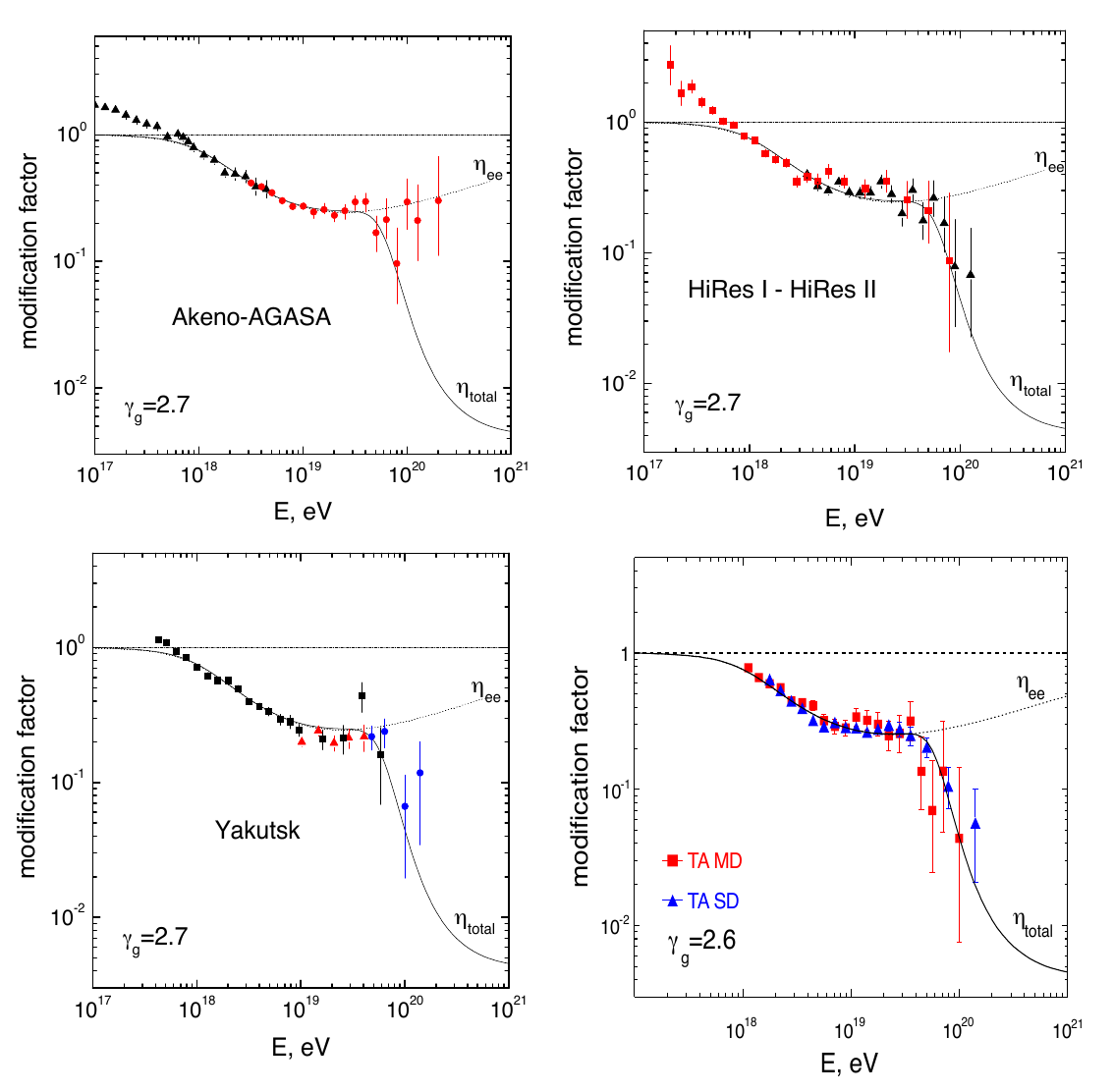}
\caption{Comparison of the modification factor with experimental data \cite{Abbasi:2002ta,Egorova:2004,Shinozaki:2006,Honda:1993,Jui:2011vm} of different experiments as labeled. Figures taken from \cite{Berezinsky:2013kfa}.}
\label{fig4}   
\end{figure}

\begin{figure}[!h]
\centering\includegraphics[scale=.8]{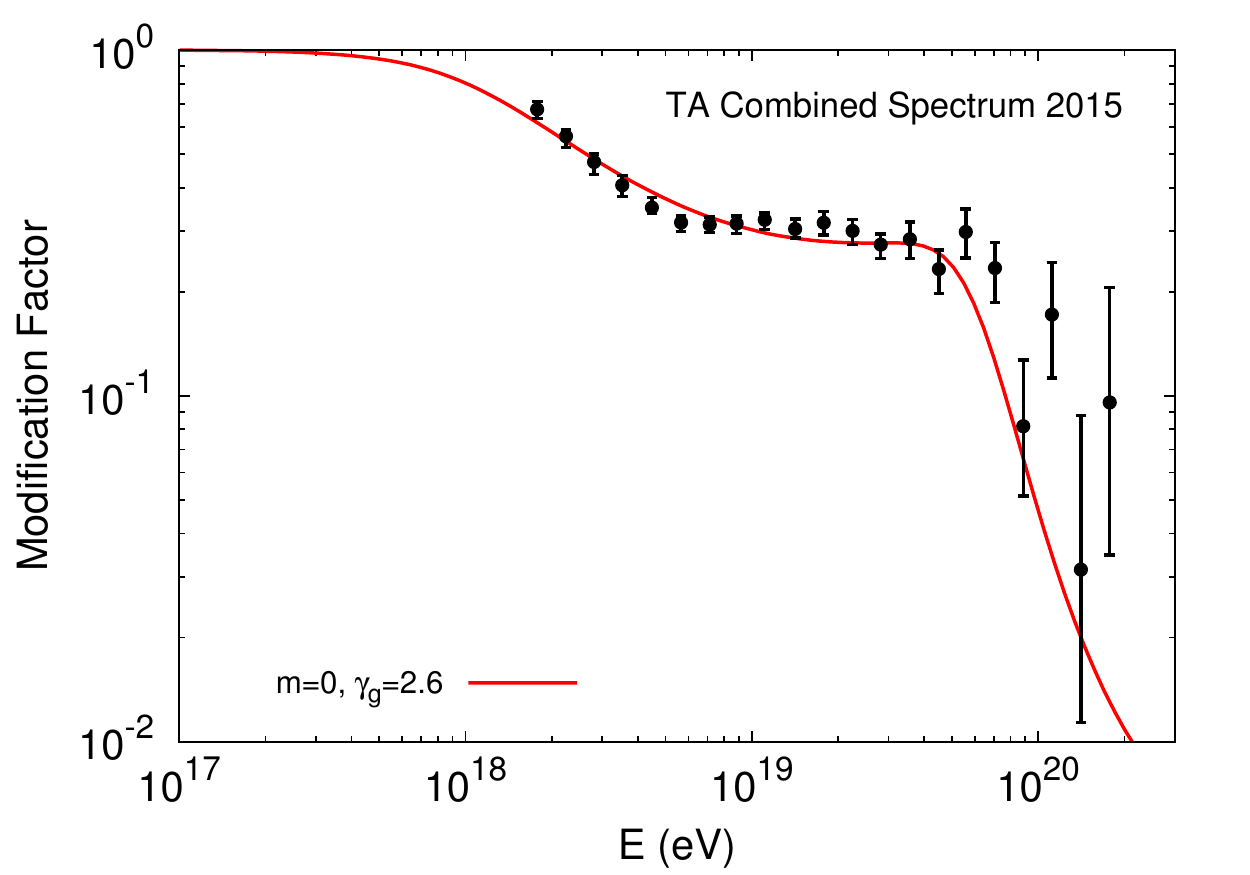}
\caption{Theoretical modification factor compared with the latest data of TA \cite{Ivanov:2015pqx}. Figure taken from \cite{Aloisio:2017kpj}.}
\label{fig4bis}   
\end{figure}

Particularly important is the behaviour associated to the pair-production energy losses that, named "dip" \cite{Berezinsky:2002nc,Aloisio:2006wv}, reproduces quite well the ankle observed in the spectrum, provided that the injection power law at the source is around $\gamma_g=2.6\div 2.7$. In figure \ref{fig4} and \ref{fig4bis} we plot the theoretical modification factor together with the experimental data of several detectors as labeled, which all claim a pure proton composition \cite{Abbasi:2002ta,Egorova:2004,Shinozaki:2006,Honda:1993,Jui:2011vm}. From these figures it is evident that the behaviour of the pair production dip reproduces quite well the ankle observed in the UHECR spectrum. 

The results presented in figures \ref{fig4} and \ref{fig4bis} refer to the case without cosmological evolution of sources, i.e. density and luminosity independent of red-shift. Assuming some kind of cosmological evolution, which typically gives a larger weight to distant sources, the transition between adiabatic and pair production energy losses arises at lower energy and the pair production dip will be deeper and slightly shifted to lower energies. As a consequence of this fact the spectral index at the sources needed to reproduce the observations will be lowered respect to the best fit value (see figure \ref{fig4} and \ref{fig4bis}) obtained without evolution \cite{Berezinsky:2002nc,Aloisio:2006wv}. 

The remarkable feature of the dip model, proposed by Berezinsky and collaborators \cite{Berezinsky:2002nc,Aloisio:2006wv,Berezinsky:2005cq,Aloisio:2007rc,DeMarco:2005ia}, is its ability of explaining experimental data with only one extra-galactic component of pure protons, directly linking the flux behaviour to the energy losses suffered by the propagating particles. 

Assuming pure protons the comparison with experimental data is restricted to the sole flux. From this analysis follows that the required emissivity depends on the power law index, which shows a best fit value that ranges from $\gamma_g=2.5$ (for strong cosmological evolution) up to $\gamma_g=2.7$ (without evolution) \cite{Berezinsky:2002nc,Aloisio:2006wv}. Using these values of $\gamma_g$ and assuming a single power law injection down to the lowest energies (GeV) results in a prohibitive energy budget for any astrophysical source. To avoid this problem, in the original papers introducing the dip model, a change in the injection spectral index was assumed \cite{Berezinsky:2002nc}: at energies below $10^{18}$ eV $\gamma_g=2.0$ while at larger energies $\gamma_g$ takes the best fit values quoted above. Under this assumption the required emissivity to reproduce UHECR data is around ${\cal L}_S=10^{45}\div 10^{46}$ erg/Mpc$^3$/yr \cite{Berezinsky:2002nc,Aloisio:2006wv,Aloisio:2015ega}.  

In figure \ref{fig4} and \ref{fig4bis} it is assumed the same maximum acceleration energy $E_{max}=10^{21}$ eV for all sources. Releasing this hypothesis and taking into account that sources can be distributed over different values of the maximum energy we can assume an injection power law index as $\gamma_g=2.0$ for all sources \cite{Kachelriess:2005xh,Aloisio:2006wv}. At energies above the minimal maximum energy, a softer spectral index arises by the convolution of the distribution function over $E_{max}$ 
 $$\frac{dN(E_{max})}{dE_{max}}\propto E_{max}^{-\alpha}$$ 
with $\alpha=1.6\div 1.7$, leading to an "effective" spectral index $\gamma_{g}^{eff}=\gamma_g+\alpha-1$ at the highest energies \cite{Kachelriess:2005xh,Aloisio:2006wv}. 

In the framework of the dip model, a suppression of the flux at low energy ($<10^{18}$ eV) is needed not only to avoid too high luminosity of the sources but also not to overshoot cosmic ray observations at energies below $10^{18}$ eV. As discussed in section \ref{sec:mag}, magnetic horizon effects can also be invoked to reduce the proton fraction at $E\le10^{18}$ eV \cite{Aloisio:2004fz} even if such effects would leave unchanged the energy budget of the sources. In the case of the dip model, therefore, the transition between galactic and extra-galactic cosmic rays will be at energies $E_{tr}< 10^{18}$ eV \cite{Aloisio:2007rc,Aloisio:2012ba}.

\subsection{Mixed composition}  
\label{sec:mix}

\begin{figure}[!h]
\centering\includegraphics[scale=.39]{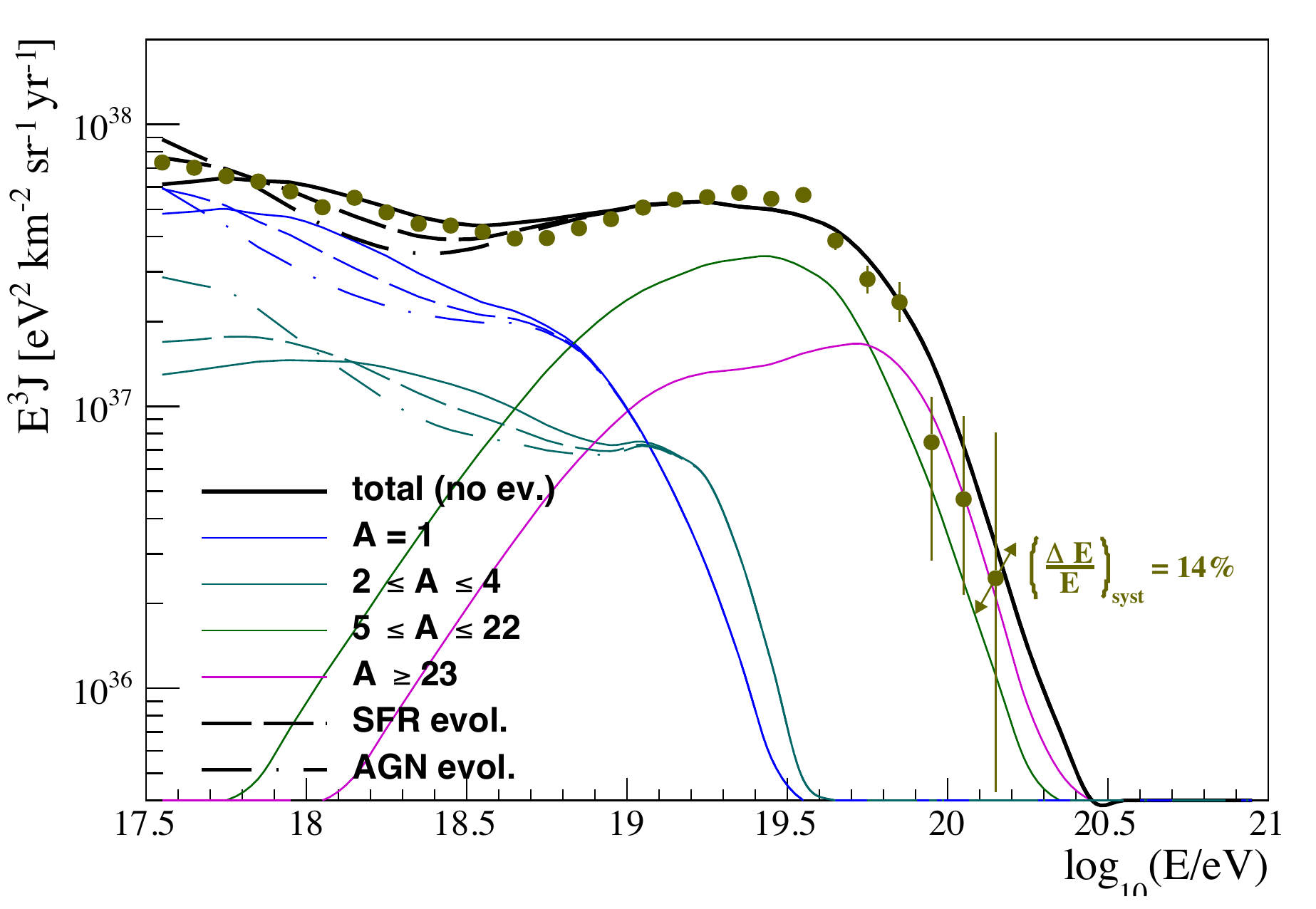}
\includegraphics[scale=.41091]{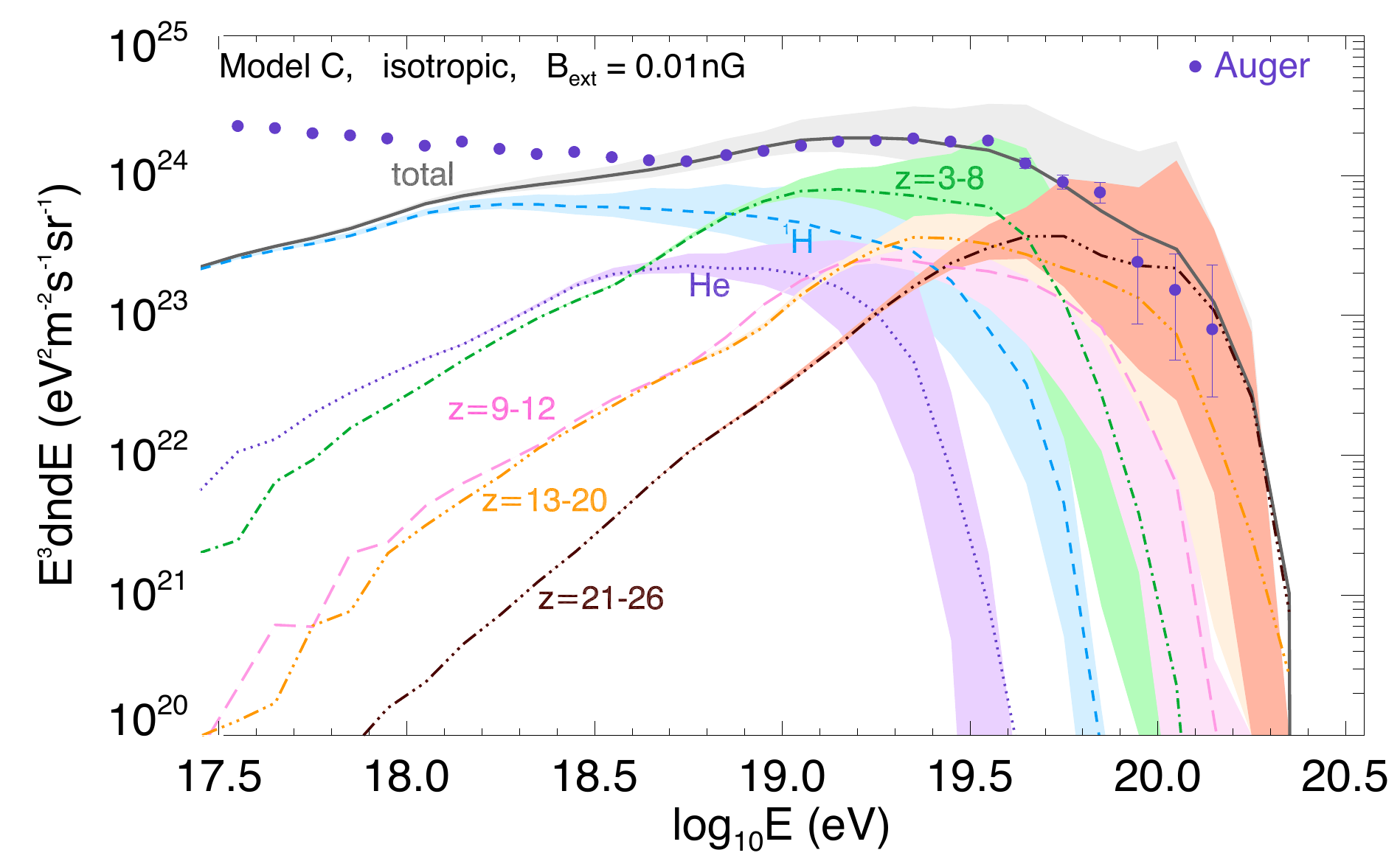}
\caption{Comparison of the Auger spectrum with theoretical expectations in the case of models with mixed composition. [Left Panel] Model with two classes of sources (figure taken from \cite{Aloisio:2015ega}). Continuos, dashed and dot-dashed lines correspond respectively to the cases of: no cosmological evolution of sources, evolution as the star formation rate and as active galactic nuclei. [Right Panel] Model with UHECR production in the internal shock of GRB (figure taken from \cite{Globus:2014fka}).}
\label{fig5}   
\end{figure}

The discussion of the previous section was centred around the hypothesis of a pure proton composition. As discussed above, a somewhat different picture arises from the Auger observations that claims a composition characterised by protons at low energies ($\le 5\times 10^{18}$ eV) and heavier nuclei at the highest. The qualitative new finding that mass composition might be mixed has served as a stimulus to build models that can potentially explain the phenomenology of Auger data. These models all show that the Auger spectrum and mass composition at $E\ge 5\times 10^{18}$ eV can be fitted at the same time only at the price of requiring very hard injection spectra for all nuclei\footnote{A somewhat different conclusion with $\gamma_g\simeq 2$ can be found assuming a negative cosmological evolution of sources \cite{Taylor:2015rla}.} ($\propto E^{-\gamma_g}$ with $\gamma_g=1\div 1.6$) and a maximum acceleration energy $E_{max}\le 5 Z\times 10^{18}$ eV \cite{Aloisio:2009sj,Aloisio:2013hya,Taylor:2013gga,Aab:2016zth,Taylor:2015rla}. The need for hard spectra can be understood taking into account that the low energy tail of the flux of UHECR reproduces the injection power law. Therefore, taking $\gamma\ge 2$ cause the low energy part of the spectrum to be polluted by heavy nuclei thereby producing a disagreement with the light composition observed at low energy.

One should appreciate here the change of paradigm that these findings imply: while in the case of a pure proton composition it is needed to find sources and acceleration mechanisms able to energise CR protons up to energies larger than $10^{20}$ eV with steep injection ($\gamma_g\simeq 2.5\div 2.7$), the Auger data require that the highest energy part of the spectrum ($E>5\times 10^{18}$ eV) has a flat injection ($\gamma_g\simeq 1.0\div 1.6$) being dominated by heavy nuclei with maximum energy not exceeding a few$\times Z\times 10^{18}$ eV  \cite{Aloisio:2013hya,Taylor:2013gga,Globus:2015xga,Unger:2015laa,Aab:2016zth}. 

By accepting the new paradigm, it follows that the Auger spectrum at energies below $5\times 10^{18}$ eV requires an additional component that, composed by protons and helium nuclei, could be, in principle, of galactic or extra-galactic origin \cite{Aloisio:2013hya,Taylor:2013gga,Globus:2015xga,Aab:2016zth}. Nevertheless, the anisotropy expected for a galactic light component extending up to $10^{18}$ eV exceeds by more than one order of magnitude the upper limit measured by Auger \cite{Abreu:2012ybu}. This observation, just restricting the analysis to Auger data, would constrain the transition between galactic and extra-galactic CR at energies below $10^{18}$ eV \cite{Giacinti:2011ww,Aloisio:2012ba}, thus excluding a galactic component at the highest energies. 

\begin{figure}[!h]
\centering
\includegraphics[scale=.40]{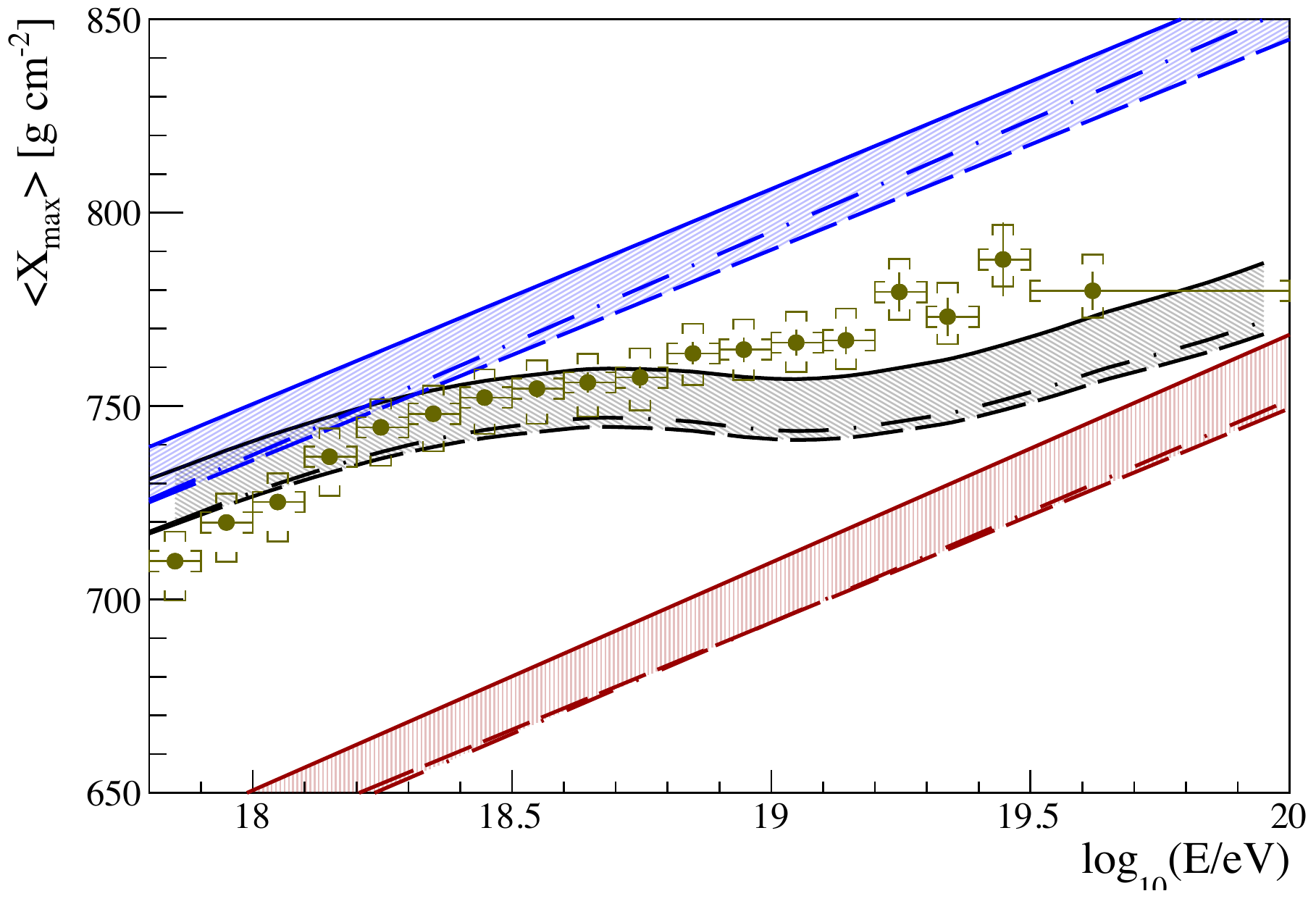}
\includegraphics[scale=.40]{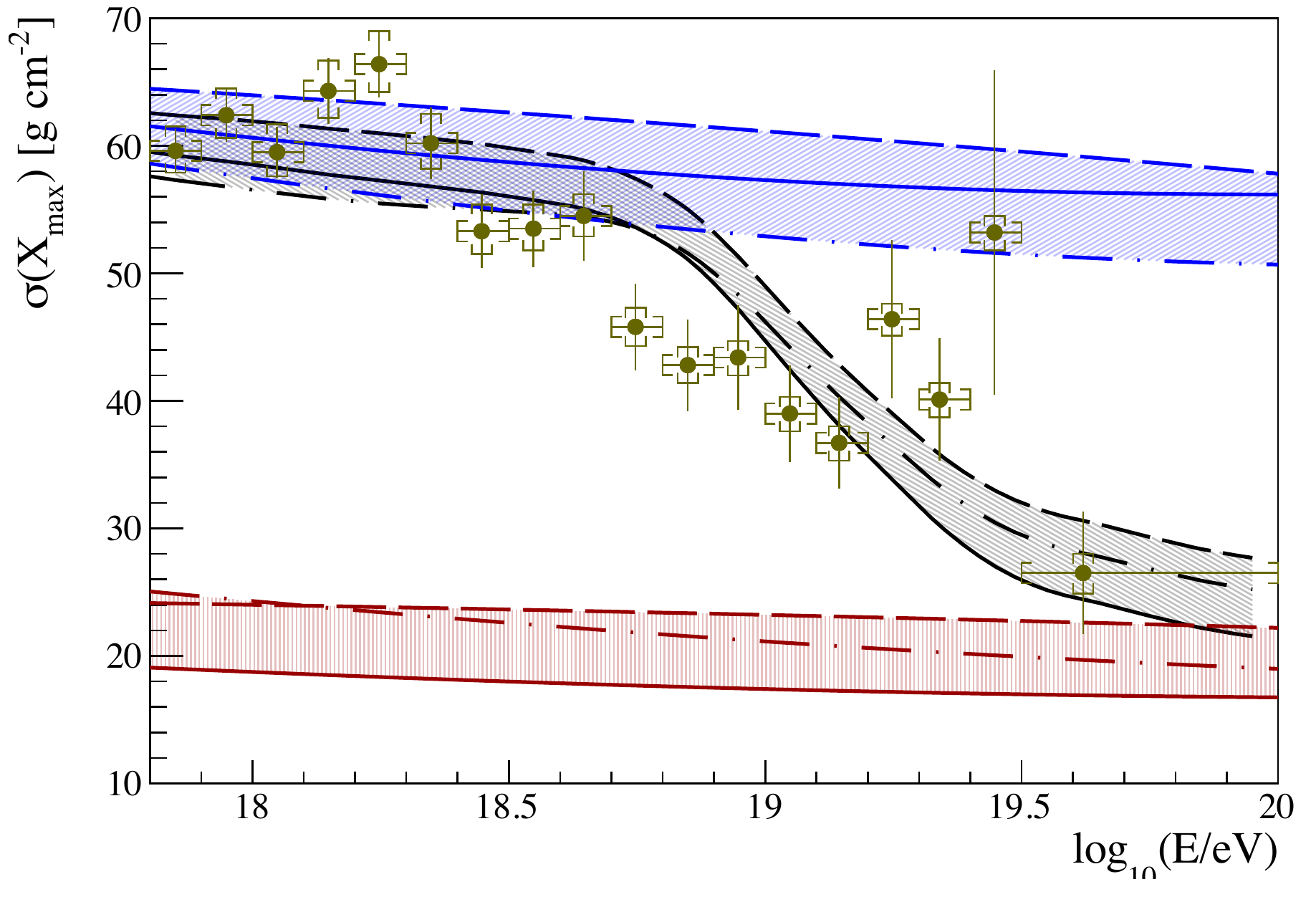}
\caption{Comparison of the elongation rate and its root mean square computed assuming the model with two classes of sources as in figure \ref{fig5} left panel (figures taken from \cite{Aloisio:2015ega}).}
\label{fig6}   
\end{figure}

\begin{figure}[!h]
\centering
\includegraphics[scale=.35]{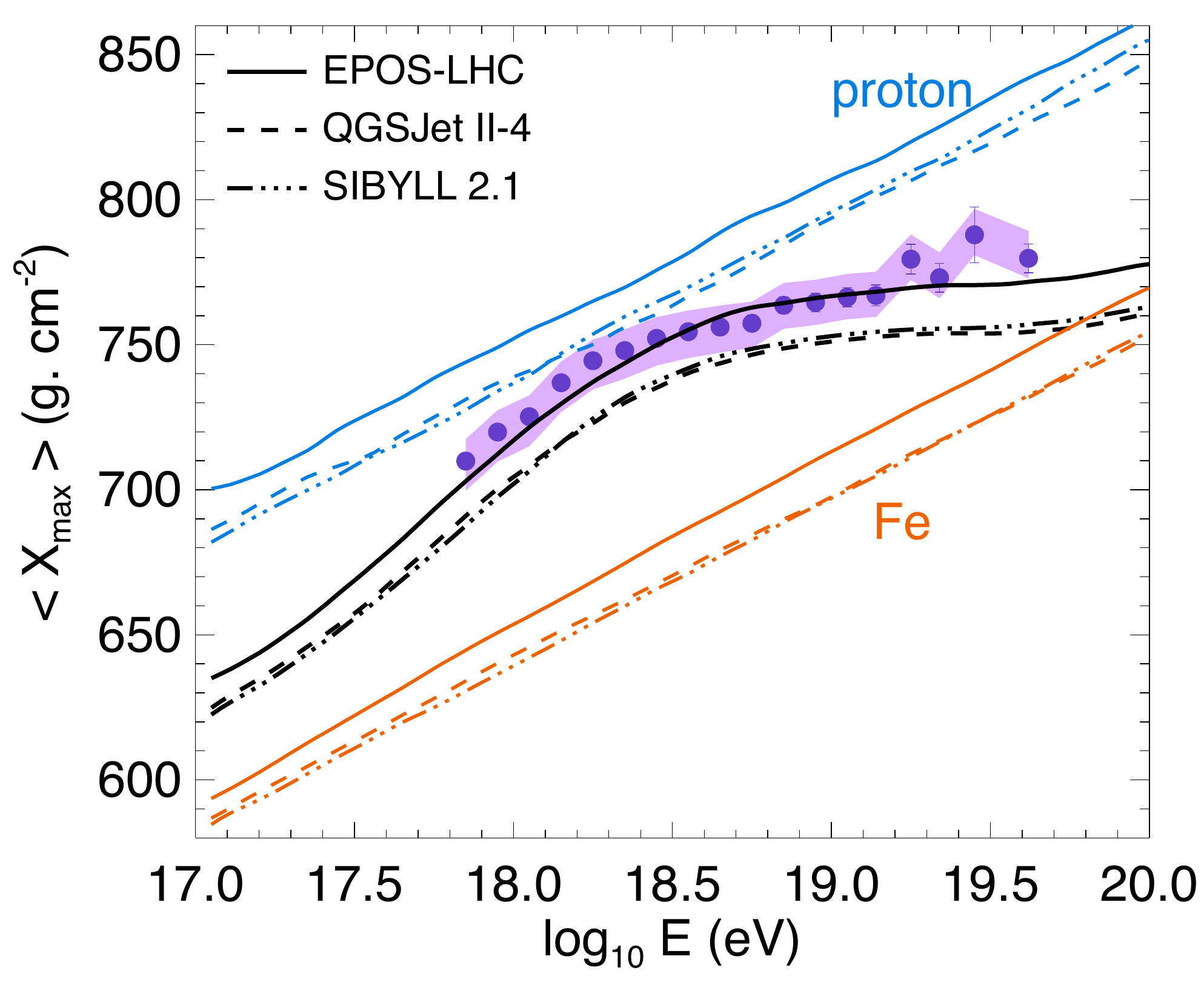}
\includegraphics[scale=.35]{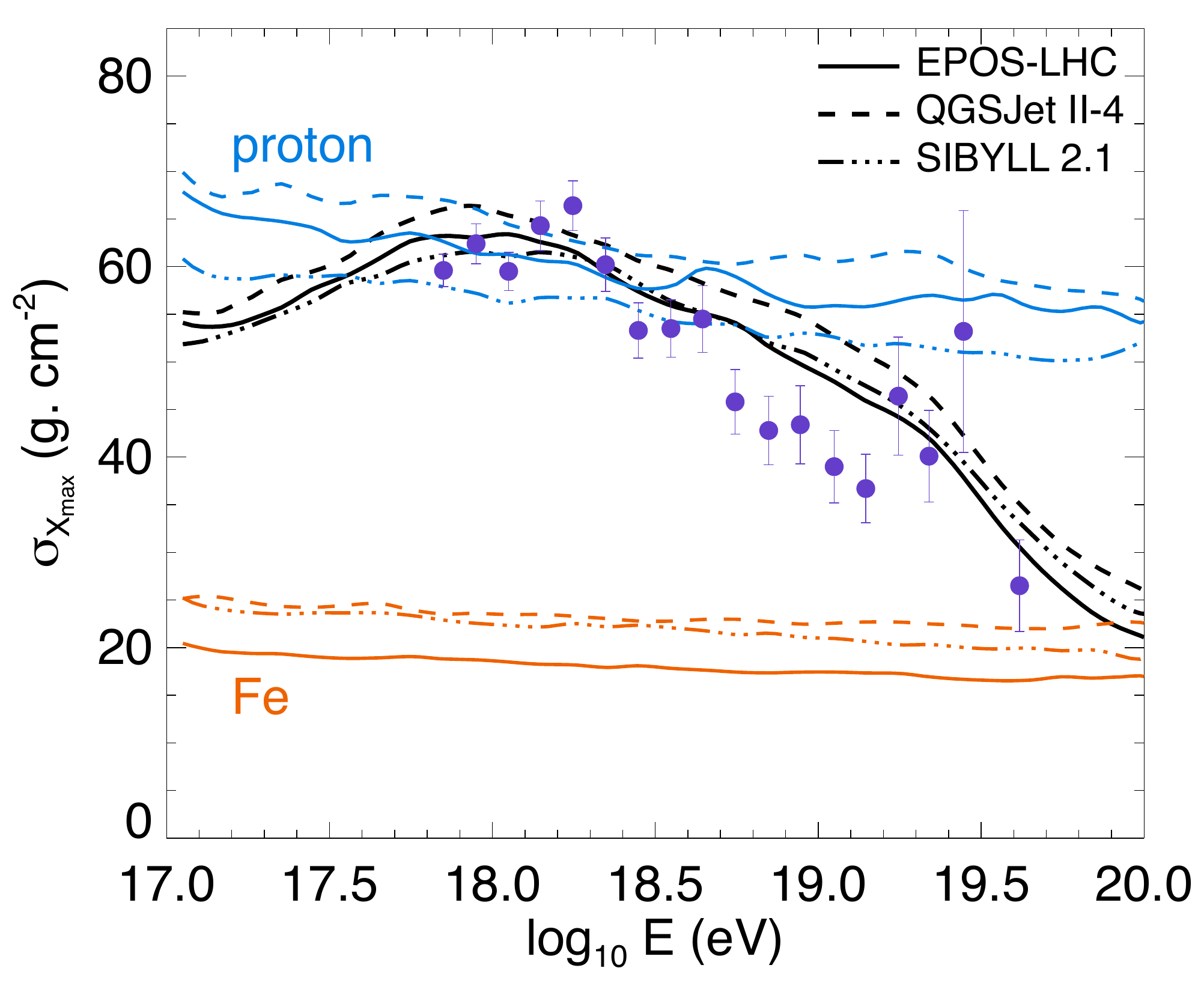}
\caption{Comparison of the elongation rate and its root mean square in the case of UHECR production in the internal shock of GRB as in figure \ref{fig5} right panel (figures taken from \cite{Globus:2014fka}).}
\label{fig7}   
\end{figure}

In order to reproduce the Auger observations, the additional (light and extra-galactic) contribution to the flux at energies below $5\times 10^{18}$ eV should exhibit a steep power law injection with $\gamma_g\simeq 2.6\div 2.7$ and a maximum acceleration energy not exceeding a few$\times 10^{18}$ eV, as for the heavier component \cite{Aloisio:2013hya,Taylor:2013gga,Globus:2015xga}. The possible origin of this radiation can be modelled essentially in two ways: (i) assuming the presence of different classes of sources: one injecting also heavy nuclei with hard spectrum and the other only proton and helium nuclei with soft spectrum \cite{Aloisio:2013hya,Taylor:2013gga} or (ii) identifying a peculiar class of sources that could provide at the same time a steep light component and a flat heavy one \cite{Globus:2014fka,Globus:2015xga,Unger:2015laa,Blasi:2015esa}. The second approach is based on a specific hypothesis on the sources that should be surrounded by an intense radiation field that, through photo-disintegration of heavy nuclei in the source neighbourhood, can provide a light component of (secondary) protons with a steep spectrum together with a hard and heavier component \cite{Globus:2014fka,Globus:2015xga,Unger:2015laa}. 

In figures \ref{fig5},\ref{fig6},\ref{fig7} we plot the comparison of the Auger data on flux and mass composition with the theoretical expectation in the case of two classes of generic sources discussed in \cite{Aloisio:2015ega} (left panel figure \ref{fig5} and figure \ref{fig6}) and in the case of a single class of sources (internal shocks in GRB) discussed in \cite{Globus:2014fka} (right panel of figure \ref{fig5} and figure \ref{fig7}). In the left panel of figure \ref{fig5} we plot also the behaviour of the spectra computed with different assumptions on the cosmological evolution of sources \cite{Aloisio:2015ega}: no cosmological evolution (solid line), the evolution typical of the star formation rate (dashed line) \cite{Gelmini:2011kg,Yuksel:2008cu,Wang:2011qc} and of active galactic nuclei (dot-dashed line) \cite{Gelmini:2011kg,Hasinger:2005sb,Ahlers:2009rf}. 

In figures \ref{fig6} and \ref{fig7}, mass composition is inferred from the mean value of the depth of shower maximum $\langle X_{max} \rangle$ and its dispersion $\sigma(X_{max})$ (computed as discussed in \cite{Abreu:2013env}). In figure \ref{fig6}, to highlight the uncertainties in the atmospheric shower development, four different models of hadronic interaction were included in the coloured bands, namely EPOS-LHC 1.99 \cite{Pierog:2006qv},  Sibyll 2.1 \cite{Ahn:2009wx}, QGSJet 01 \cite{Kalmykov:1997te} and  QGSJet 02 \cite{Ostapchenko:2005nj}. In figure \ref{fig7} different lines correspond to different hadronic interaction models as labeled. 

\subsection{Secondary cosmogenic messengers} 
\label{sec:sec}

The propagation of UHECR in the intergalactic space, through the interactions with CMB and EBL, gives rise to the production of several unstable particles that in turn produce high energy photons, electrons/positrons and neutrinos. The possible detection of these signal carriers, soon realised after the proposal of the GZK cut-off \cite{Stecker:1968uc,Beresinsky:1969qj,Stecker:1973sy,Berezinsky:1975zz}, is extremely important to constrain models for UHECR sources, composition and the details of propagation \cite{Engel:2001hd,Kalashev:2002kx,Hooper:2004jc,Seckel:2005cm,DeMarco:2005kt,Allard:2006mv,Anchordoqui:2007fi,Takami:2007pp,Ahlers:2009rf,Kotera:2010yn,Berezinsky:2010xa,Stanev:2014asa,Aloisio:2015ega,Heinze:2015hhp,Berezinsky:2016feh}.

\subsubsection{Neutrinos}
\label{sec:nu}

There are two processes by which neutrinos can be emitted in the propagation of UHECR: (i) the decay of charged pions, produced by photo-pion production, $\pi^{\pm}\to \mu^{\pm}+\nu_{\mu}(\bar{\nu}_{\mu})$ and the subsequent muon decay $\mu^{\pm}\to e^{\pm}+\bar{\nu}_{\mu}(\nu_{\mu})+\nu_e(\bar{\nu}_e)$; (ii) the beta-decay of neutrons and nuclei produced by photo-disintegration: $n\to p+e^{-}+\bar{\nu}_e$, $(A,Z)\to (A,Z-1)+e^{+}+\nu_e$, or $(A,Z)\to (A,Z+1)+e^{-}+\bar{\nu}_e$. These processes produce neutrinos in different energy ranges: in the former the energy of each neutrino is around a few percent of that of the parent nucleon, whereas in the latter it is less than one part per thousand (in the case of neutron decay, larger for certain unstable nuclei). This means that in the interaction with CMB photons, which has a threshold Lorentz factor of about $\Gamma\ge 10^{10}$, neutrinos are produced with energies of the order of $10^{18}$~eV and $10^{16}$~eV respectively. Interactions with EBL photons contribute, with a lower probability than CMB photons, to the production of neutrinos with energies of the order of $10^{15}$~eV in the case of photo-pion production and $10^{14}$~eV in the case of neutron decay (see \cite{Aloisio:2015ega} and reference therein).   

\begin{figure}[!h]
\centering
\includegraphics[scale=.40]{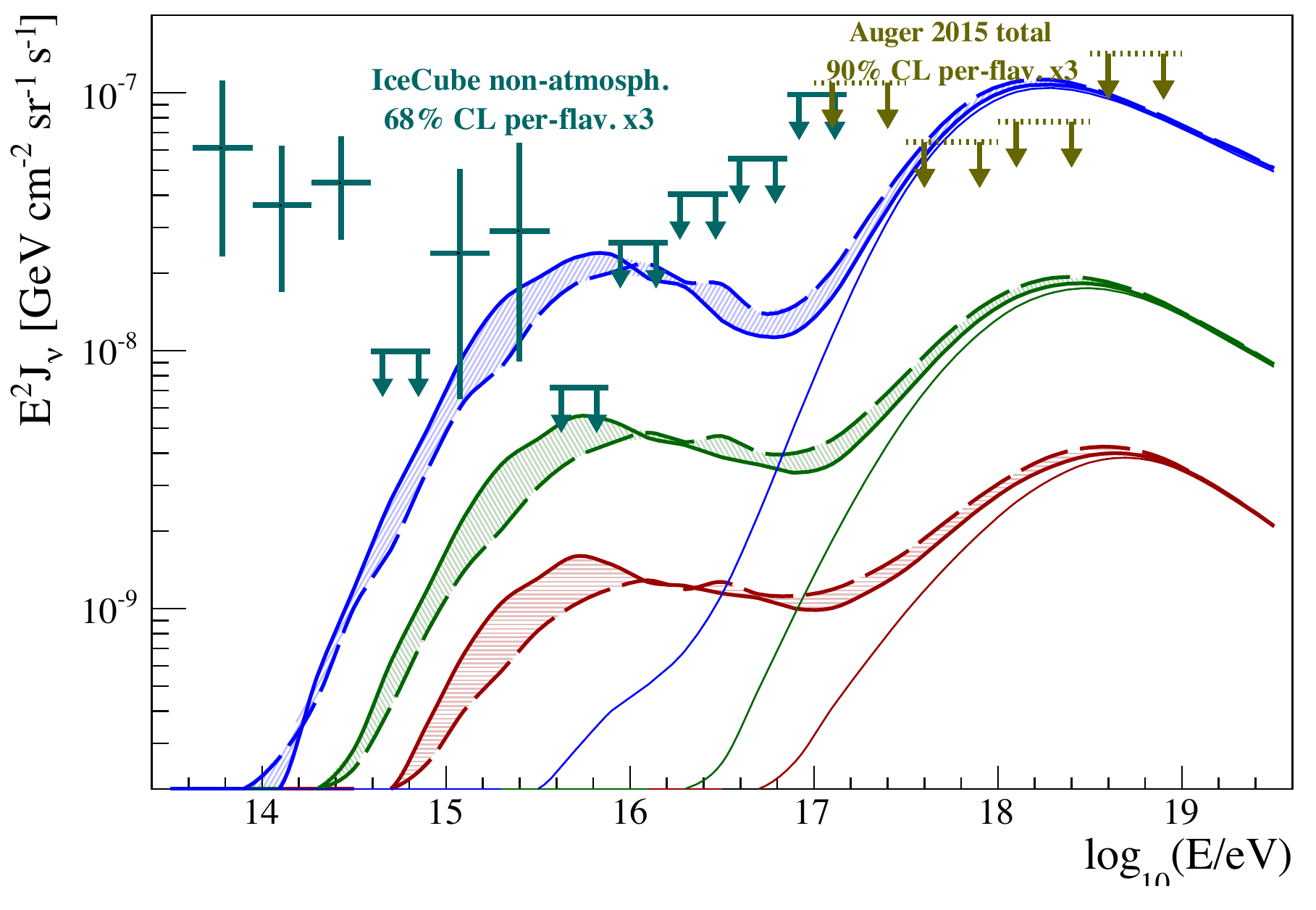}
\includegraphics[scale=.40]{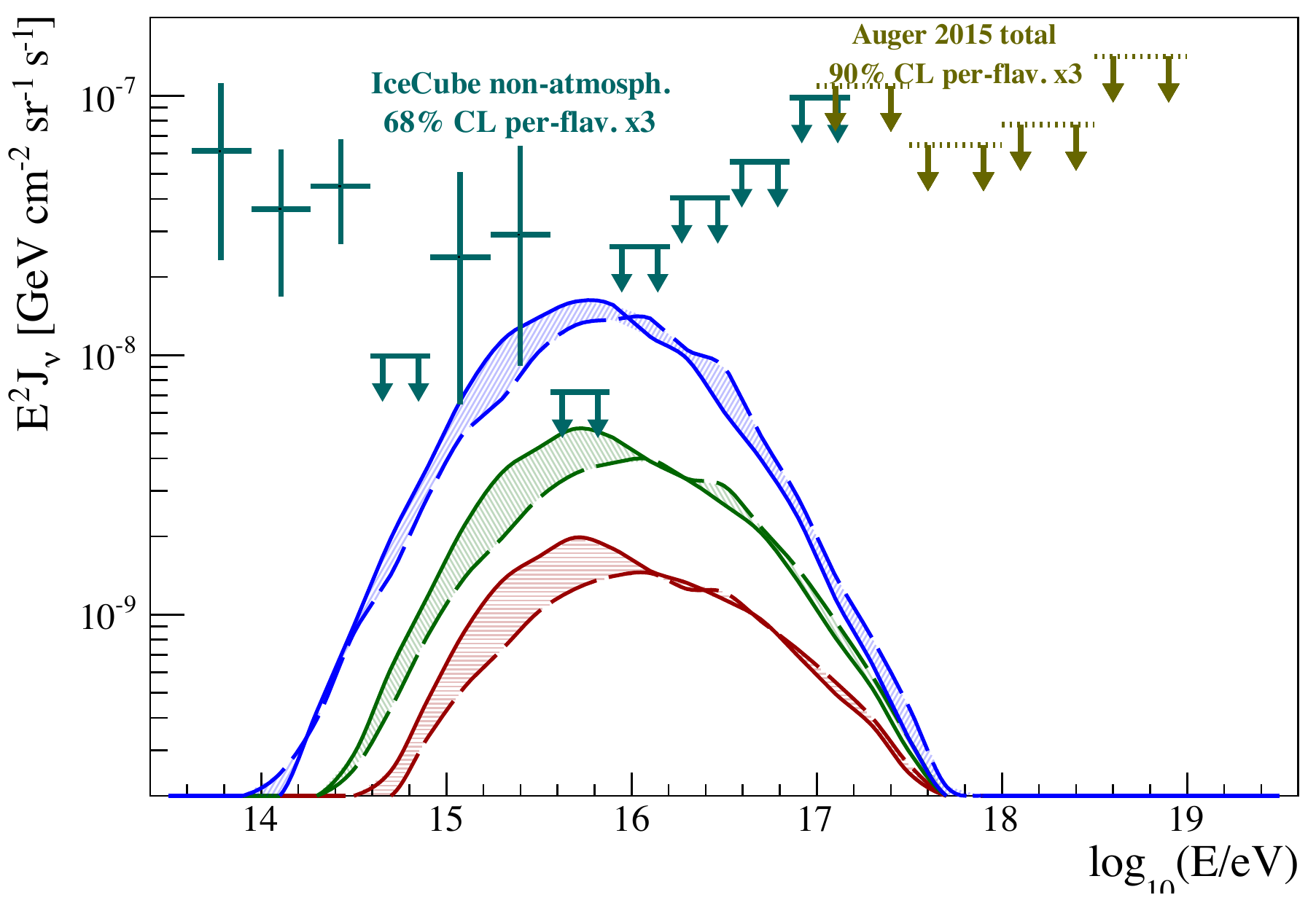}
\caption{[Left Panel] Fluxes of neutrinos in the case of the dip model. The three different fluxes correspond to different assumptions on the cosmological evolution of sources (from bottom to top): no evolution (red), SFR (green) and AGN (blue), coloured bands show the uncertainties due to the EBL model \cite{Stecker:2005qs,Stecker:2006eh,Kneiske:2003tx}. Thin solid lines are neutrino fluxes obtained taking into account the sole CMB field. [Right Panel] Neutrino fluxes in the case of mixed composition (as shown in figure \ref{fig5} left panel) with the same color code of left panel. Experimental points are the observation of IceCube on extra-terrestrial neutrinos \cite{Aartsen:2013jdh,Aartsen:2013bka} and the Auger limits on neutrino fluxes \cite{Abreu:2013zbq}. Figures taken from \cite{Aloisio:2015ega}.} 
\label{fig8}   
\end{figure}

The flux of secondary neutrinos is very much sensitive to the composition of UHECR. In figure \ref{fig8} we plot the flux of cosmogenic neutrinos expected in the case of the dip model (left panel) and in the case of mixed composition (right panel). Comparing the two panels of figure \ref{fig8} it is evident the huge impact of the composition on the expected neutrino flux: heavy nuclei provide a reduced flux of neutrinos because the photo-pion production process in this case is subdominant. 

The production of cosmogenic neutrinos is almost independent of the variations in sources' distribution because the overall universe, up to the maximum red-shift, could contribute to the flux. Typically, the maximum red-shift of astrophysical structures is expected around $z_{max}\simeq 10$, which is the redshift of the first stars (pop III) \cite{Berezinsky:2011bb}. Once produced at these cosmological distances neutrinos travel toward the observer almost freely, except for the adiabatic energy losses and flavour oscillations, the opacity of the universe to neutrinos being relevant only at redshifts $z \gg 10$ \cite{Weiler:1982qy,Gondolo:1991rn}. This is an important point that makes neutrinos a viable probe not only of the mass composition of UHECR but also of the cosmological evolution of sources. In figure \ref{fig8} three different hypothesis on the cosmological evolution of sources are taken into account: no cosmological evolution (red bands), evolution typical of the star formation rate (green band) \cite{Gelmini:2011kg,Yuksel:2008cu,Wang:2011qc} and of active galactic nuclei (blue band) \cite{Gelmini:2011kg,Hasinger:2005sb,Ahlers:2009rf}. 
 
There is a solid consensus about the light composition of UHECR in the low energy part of the observed spectrum. This assures a flux of cosmogenic neutrinos in the PeV energy region, produced by the protons' photo-pion production on the EBL photons. Coloured bands in figure \ref{fig8} show the uncertainties connected with the EBL background \cite{Stecker:2005qs,Stecker:2006eh,Kneiske:2003tx}. Another important uncertainty in the expected neutrino flux comes from the contribution of UHECR sources at high red-shift. Given the energy losses suffered by UHE protons and nuclei, sources at red-shift larger than $z>1$ can be observed only in terms of cosmogenic neutrinos \cite{Aloisio:2015ega,Heinze:2015hhp,Berezinsky:2016feh}. Therefore a lack in the UHE neutrino flux could also be accommodated invoking a lack of sources at high red-shift.  

\subsubsection{Gamma rays}
\label{sec:gamma}

While neutrinos reach the observer without being absorbed, high energy photons and electrons/positrons colliding with astrophysical photon backgrounds (CMB and EBL) produce electromagnetic cascades (EMC) through the processes of pair production (PP, $\gamma+\gamma_{CMB,EBL}\to e^{+}+e^{-}$) and Inverse Compton Scattering (ICS, $e+\gamma_{CMB,EBL}\to \gamma + e$). While PP is characterised by a threshold the ICS process does not. From this simple observation follows that once a cascade is started by a primary photon/electron/positron it develops since the energy of photons produced by ICS are still above the PP threshold. The final output of the cascade, i.e. what is left behind when the cascade is completely developed, is a flux of low energy photons all with energies below the PP threshold. 

The two astrophysical backgrounds CMB and EBL against which the EMC develops are characterised by typical energies $\epsilon_{CMB}\simeq 10^{-3}$ eV and $\epsilon_{EBL}\simeq 1$ eV. Hence, the typical threshold energy scale for pair-production will be respectively\footnote{Numerical values quoted here should be intended as reference values being background photons distributed over energy and not monochromatic.} ${\mathcal E}_{CMB}=m_e^2/\epsilon_{CMB}=2.5\times 10^{14}$ eV and ${\mathcal E}_{EBL}=m_e^2/\epsilon_{EBL}=2.5\times 10^{11}$ eV. The radiation left behind by the cascade will be restricted to energies below ${\mathcal E}_{EBL}$.

\begin{figure}[!h]
\centering
\includegraphics[scale=.40]{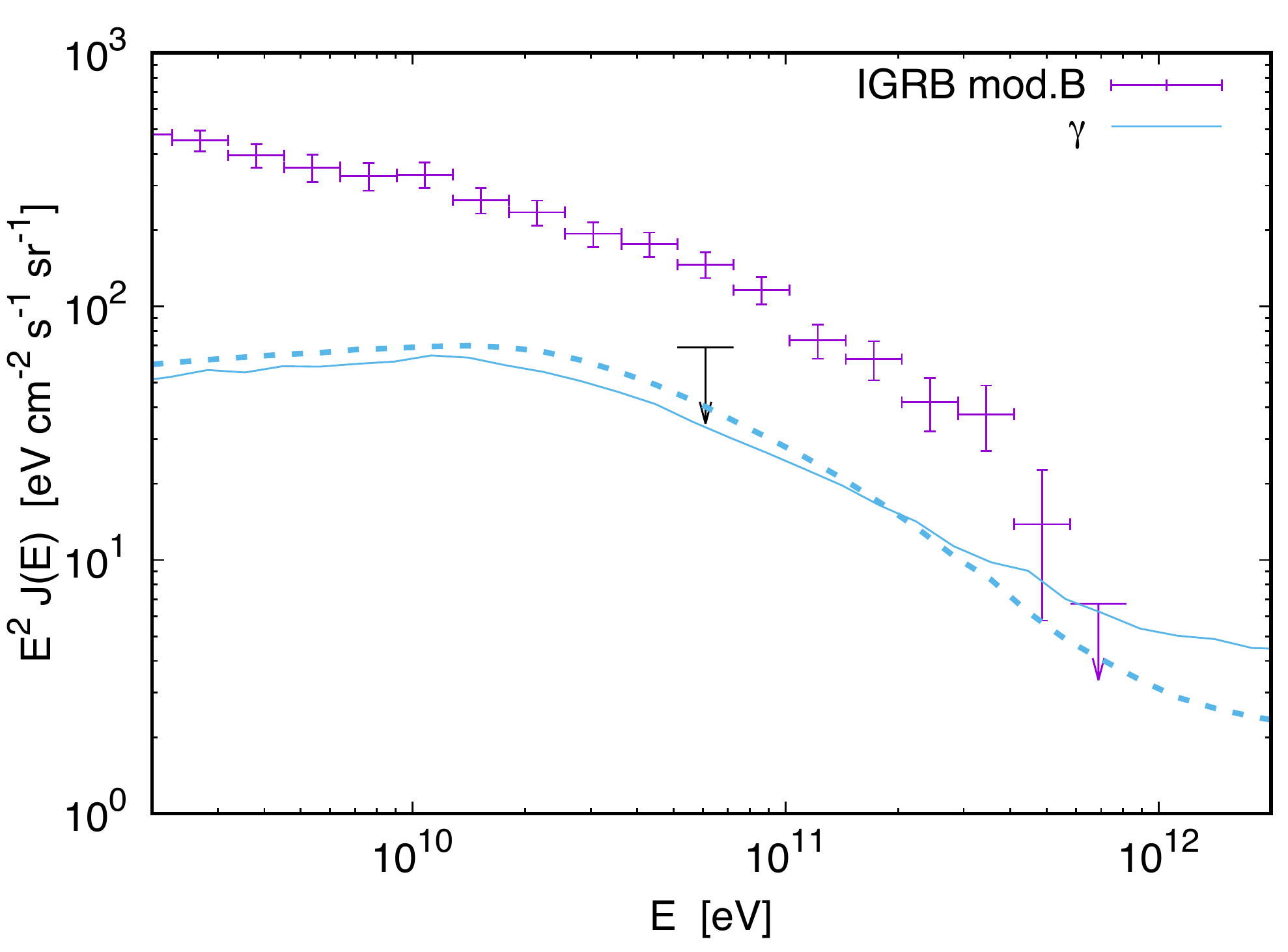}
\includegraphics[scale=.35]{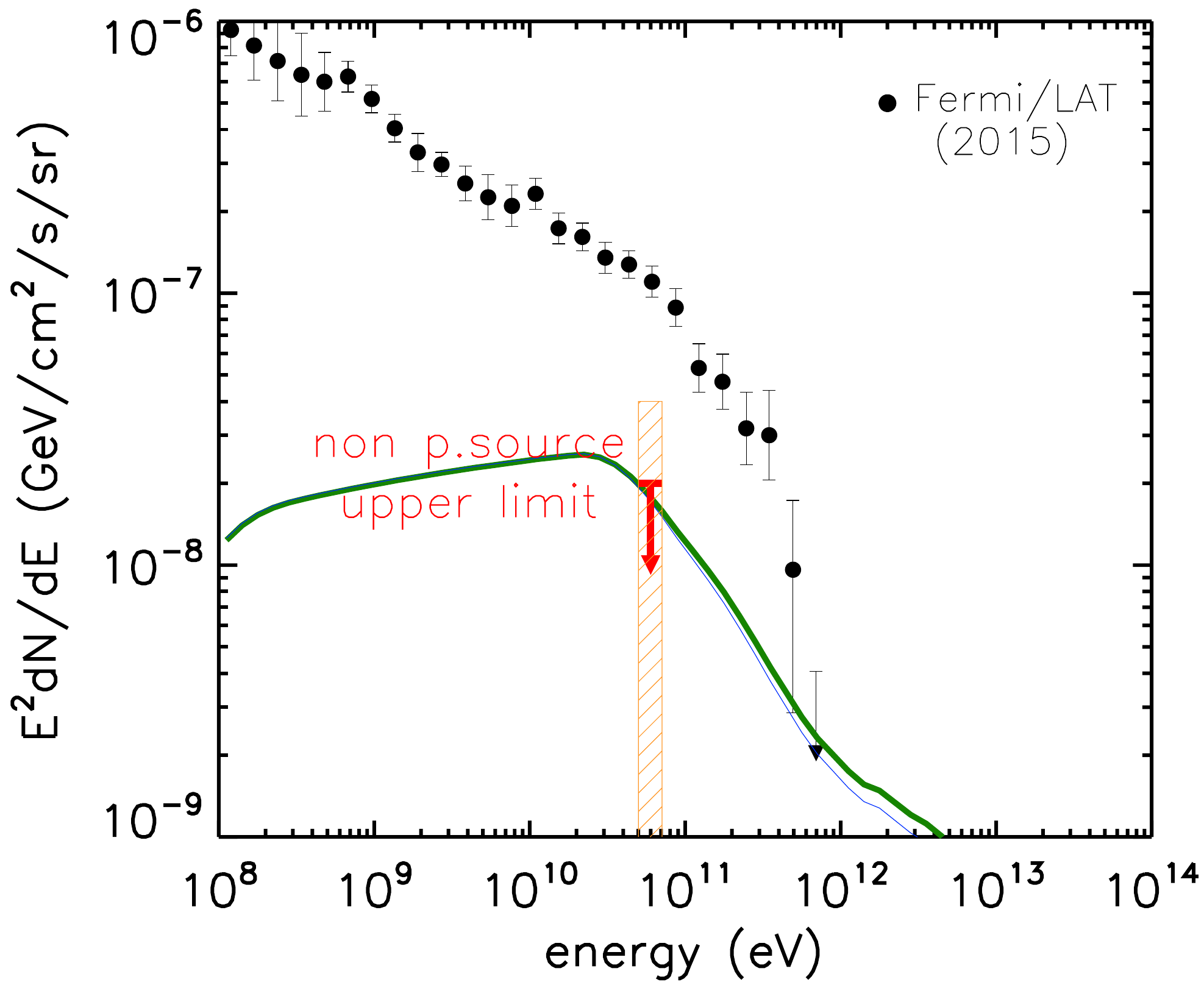}
\caption{Spectra of cosmogenic gamma rays obtained in the case of pure proton composition of UHECR without cosmological evolution of sources, as computed in \cite{Berezinsky:2016jys} (left panel) and in \cite{Liu:2016brs} (right panel), together with the Fermi-LAT data on diffuse gamma ray background, as in model-B (left panel) and model-A (right panel) of the analysis presented in \cite{TheFermi-LAT:2015ykq}.} 
\label{fig8bis}   
\end{figure}

The cascade development has a universal nature independent of the energy and spectrum of the initial photon/pair. It can be proved\footnote{For a recent detailed discussion of EMC development on CMB and EBL see \cite{Berezinsky:2016feh} and references therein.} that the spectrum of photons produced in the cascade, those left behind with energy below threshold, is always of the type: 
\begin{equation}
n_{\gamma}(E_{\gamma})\propto \left \{ \begin{array}{ll}
E_{\gamma}^{-3/2} & E_{\gamma} < {\mathcal E}_X \\
&\\
E_{\gamma}^{-2} & {\mathcal E}_X \le E_{\gamma} \le {\mathcal E}_{EBL}
\end{array}
\right.
\label{eq:EMC-spectrum}
\end{equation} 
being ${\mathcal E}_X=(1/3) {\mathcal E}_{EBL} \epsilon_{CMB}/\epsilon_{EBL}$ the (average) minimum energy of a photon produced through the ICS mechanism by an electron with the minimum allowed energy ${\mathcal E}_{EBL}/2$ \cite{Blumenthal:1970nn,Blumenthal:1970gc,Ginzburg:1990sk}. The normalisation of the spectrum (\ref{eq:EMC-spectrum}) can be easily determined imposing energy conservation, i.e. the total energy of the cascading photons should correspond to the energy of the photon/pair that started the cascade. 

In the case of an expanding universe the cascade acquires some dependence on the energy and spectrum of the initiating particle. A cascade started at certain red-shift $z$ still develops independently of energy and spectrum of the primary photon/pair, nevertheless the spectrum of photons observed at $z=0$ will depend on the red-shift $z$ where the cascade initiated \cite{Berezinsky:2016feh,Ginzburg:1990sk}. This is a direct consequence of the cosmological evolution of astrophysical backgrounds that implies an increasing energy and density at increasing red-shift.

The propagation of UHECR in astrophysical backgrounds certainly produces EMC started by pairs and photons coming by pair-production and photopion production of UHECR on CMB and EBL. These cascades transform the energy lost by UHECR in low energy gamma ray photons, with the characteristics discussed above (spectrum as in (\ref{eq:EMC-spectrum}) in the energy range $100 ~{\rm MeV}<E<100 ~{\rm GeV}$), that in turn contribute to the diffuse gamma ray background \cite{Beresinsky:1969qj,Berezinsky:1975zz,Ferrigno:2004am,Berezinsky:2010xa,Ahlers:2010fw,Gelmini:2011kg,Berezinsky:2016feh,Liu:2016brs}. Therefore, the observation of a diffuse extra-galactic gamma ray background by the Fermi-LAT satellite \cite{Abdo:2010nz,Ackermann:2014usa} can be used to constrain models of UHECR. The observed fast decrease in energy ($\propto E^{-2.4}$) of the diffuse background already limits models of pure proton composition, which maximise the production of secondary cosmogenic particles. Versions of the dip model (see section \ref{sec:dip}) with strong red-shift evolution of sources seem already ruled out by Fermi-LAT observations \cite{Berezinsky:2010xa}. These constrains could become even more severe taking into account a recent analysis of the Fermi-LAT collaboration \cite{TheFermi-LAT:2015ykq}, where different models for unresolved gamma ray sources are taken into account to determine the 'true' diffuse extra-galactic gamma ray background. 

In figure \ref{fig8bis} (left panel taken from \cite{Berezinsky:2016jys} and right panel from \cite{Liu:2016brs}) we plot the expected gamma ray background in the case of the dip model without cosmological evolution of sources in comparison with the experimental data of Fermi-LAT with two models for the unresolved gamma ray sources as discussed in \cite{TheFermi-LAT:2015ykq}. 

The detection of point-like gamma ray sources at GeV-TeV energies could also be a promising way to reveal powerful astrophysical accelerators of UHECR \cite{Ferrigno:2004am,Essey:2009ju,Essey:2010er}. During their journey from the source to the observer UHECR could give rise to EMC cascades that, along the line of sight of the source, will shine as a point-like source in the gamma ray band. This possibility critically depends on the magnetic field in the intergalactic space as it could modify the spatial development of the cascade. Therefore, to reveal a point-like source of this kind it is needed to confine the cascade within small angular size around the source line of sight and the corresponding magnetic field should be quite low ($B<10^{-5}$ nG) \cite{Ferrigno:2004am,Essey:2009ju,Essey:2010er}. 

Let us conclude stressing the importance of the magnetic field in the physics of EMC. Cascades can be sustained only if the process of ICS dominates over electron interactions. Increasing the magnetic field synchrotron interaction becomes more and more important with the net result of producing low energy ($\le$MeV) photons thus damping the cascade development. 

\section{Astrophysical sources} 
\label{sec:source}

\begin{figure}[!h]
\centering
\includegraphics[scale=.50]{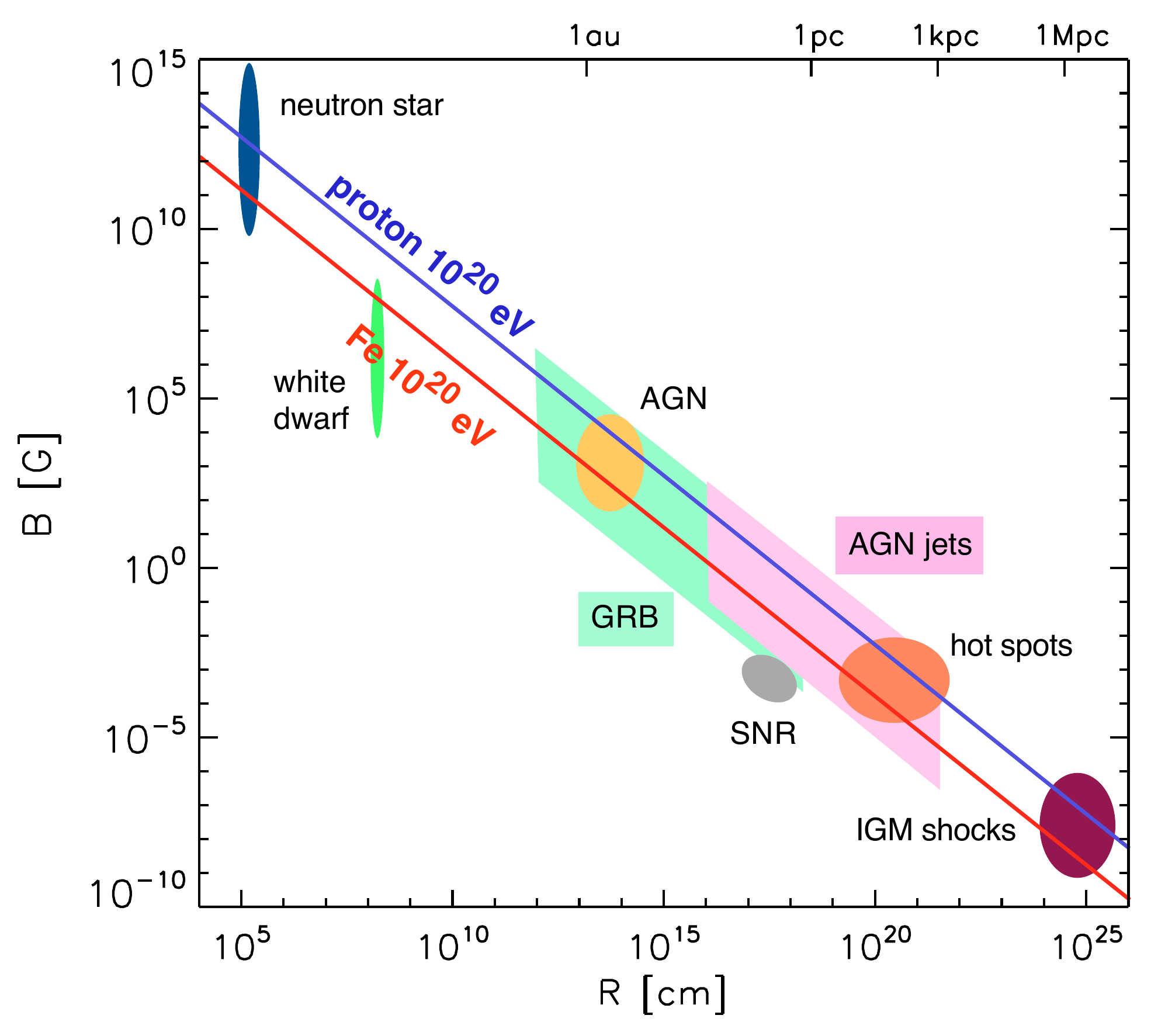}
\caption{Hillas plot \cite{Hillas:1985is} obtained by imposing the condition $r_L(E)<R$, the two solid lines correspond to protons and iron nuclei with $E=10^{20}$ eV as labeled. Also shown are several astrophysical objects that meet the requirements of size/magnetic field needed for the acceleration process. Figure taken from \cite{Kotera:2011cp}.}
\label{fig:HillasPlot}   
\end{figure}

We do not know what the sources of UHECR are. Therefore it is important to define general requirements that a hypothetic accelerator should fulfil in order to reach the extreme energies observed. If the size of the accelerator is $R$, a necessary condition to reach the energy $E$ is that particles with this energy should remain trapped inside the source, hence the Larmor radius of the particle should be: $r_L(E)<R$. This condition fixes a relation between the size of the accelerator $R$ and the magnetic field $B$ in the acceleration region that is at the base of the so-called Hillas plot. In figure \ref{fig:HillasPlot} we show the Hillas plot \cite{Hillas:1985is} with the curves relative to the condition $r_L<R$ (see Eq. (\ref{eq:rL})), the corresponding energy of the accelerated particles, and several astrophysical objects that match this condition. It is interesting to note here that the acceleration of nuclei, thanks to larger electric charge, is less demanding and can be easier achieved respect to the case of protons. 

Following \cite{Waxman:2005id,Blasi:2012yx} the general idea at the base of the Hillas plot can be further refined by assuming that the accelerator moves (as for shocks) with either relativistic or non-relativistic velocity. Let us consider first the non-relativistic case. The condition $r_L(E)<R$ on the magnetic field can be easily transformed in a condition on the magnetic energy density $\epsilon_B=B^2/4\pi$. This quantity should be lower that the total ram pressure $\epsilon_B<\rho V^2$ and this fixes a lower limit on the luminosity of the source: 
\begin{equation} 
L=4\pi R^2 V \frac{\rho V^2}{2} >2\pi R^2 V \epsilon_B \simeq 1.6\times 10^{45} Z^{-2} \left (\frac{E}{10^{20}{\rm eV}}\right )^2 \beta ~{\rm erg/s}
\label{eq:L_limit}
\end{equation}
being $Z$ the electric charge of the particle and $\beta=V/c$ the velocity of the accelerator. As discussed in \cite{Blasi:2012yx}, the condition (\ref{eq:L_limit}) is somewhat uncertain in the scaling with $\beta$, as it depends on the details of particles transport in the acceleration region. 

In the case of a relativistic motion of the acceleration region, i.e. with a Lorentz factor $\Gamma \gg 1$, it is useful to introduce the co-moving reference frame, all quantities in this frame will be indicated with a prime. The condition discussed above for acceleration in the co-moving frame becomes: $r_L(E')<R'=R/\Gamma$, using the Lorentz contraction of length. Since the energy density transforms as $\Gamma^2$, we can rewrite the condition on luminosity in the case of a relativistic motion of the accelerator as: 
\begin{equation}
L>4\pi R^2c\Gamma^2\epsilon_{B'}\simeq 10^{47}\Gamma^2 Z^{-2} \left (\frac{E}{10^{20} eV}\right )^2 ~{\rm erg/s}~.
\label{eq:L_lim_rel}
\end{equation}

What is more relevant in Eq. (\ref{eq:L_limit}) and (\ref{eq:L_lim_rel}) is the strong dependence on the electric charge of the particle $Z^{-2}$ that softens the limit in the case of nuclei.

Another quantity that can be used to constrain sources of UHECR is the energy input per unit volume and time, i.e. the sources emissivity. The flux of UHECR can be roughly estimated as: 

\begin{equation}
J(E)\simeq \frac{c}{4\pi} Q(E)\tau_{loss}(E)
\label{eq:flux}
\end{equation}
where $Q$ is the injection rate per unit volume and $\tau_{loss}$ the time scale of energy losses. Comparing Eq. (\ref{eq:flux}) with the observed flux we can immediately deduce, at fixed energy, the expected emissivity of the sources around $3\times 10^{45}$ erg/Mpc$^3$/yr at $10^{19}$ eV. This quantity can be compared with the known emissivity of well defined sources. For instance, AGN typically show bolometric luminosities in gamma rays in the range between $L_{bol}\simeq 10^{43}$ erg/s, for Seyfert galaxies and radio-quite quasars, and $L_{bol}\simeq 10^{47}$ erg/s, for radio-loud quasars, with typical number density in the range $10^{-5}\div 10^{-4}$ Mpc$^{-3}$ \cite{Woo:2002un,Jiang:2006vv}. Therefore AGN would meet the energy requirements if they emit a fraction in the range $10^{-4}\div 10^{-3}$ of their bolometric luminosity in UHECR. 

In the following we will address three general categories of possible astrophysical sources distinguishing among the acceleration mechanisms at work: relativistic and non-relativistic shocks and strong electric fields, as those produced by rotating magnetised stars. 

\subsection{Non relativistic shocks and large scale structures} 
\label{sec:non_rel_acc} 

There are not many examples of non-relativistic shocks able to accelerate particles at the extreme energies of UHECR. The most noteworthy case is certainly represented by shocks produced during the formation of clusters of galaxies \cite{Kang:1995xw,Kang:2012nz,Hong:2014yda}. Generally speaking, the formation of large scale structures in the universe naturally leads to supersonic motion of plasma that, fuelled by gravitational forces, gives rise to shock waves. These shocks can be formed by clusters merging as well as in the case of accretion of an isolated cluster, in this case a pre-existing gravitational potential attracts the accretion of gas from the outer space. This is a very important mechanism through which filaments, that connect clusters in the cosmic web, are formed. 

In filaments typical accretion velocities reached are at the level of $10^3\div 10^4$ Km/s with a background temperature relatively small at the level of $10^{5}$ $^\circ$K, hence the shock waves that can be formed will be relatively strong with values of the Mach number at the level of $M_s\simeq 10\div 100$. On the other hand, shocks produced by clusters merging are weaker ($M_s\ll10$) because formed in the intra-cluster medium with a typical temperature at the level of $10^8$ $^\circ$K, being clusters already virialised structures \cite{Gabici:2002ie}. In both cases of accretion and merging shocks the available luminosity is typically of the order of $L\simeq 10^{45}$ erg/s therefore compatible with the estimates discussed above.

This kind of accelerators operate for very long time, of the order of the age of the universe, and the accelerated particles are typically confined inside the cluster volume \cite{Berezinsky:1996wx}. The maximum energy attainable with these mechanisms is constrained by the magnetic field at the shock, which fixes the acceleration time, and by the energy losses of particles. Particularly relevant are losses due to photo-pion production for protons and photo-disintegration for nuclei that fix the maximum energy at the level of $5\div 10 \times 10^{19}$ eV \cite{Kang:1995xw,Kang:2012nz,Hong:2014yda}. 

Shocks in clusters of galaxies, also called cosmological shocks, may represent an option for UHECR acceleration only in the case of no substantial flux observed at energies higher than a few$\times 10^{19}$ eV. The spectrum of the accelerated particles produced by cosmological shocks can be determined applying the general theory of particle acceleration at non-relativistic shocks, for a recent review see \cite{Blasi:2013rva}.

\subsection{Relativistic shocks} 
\label{sec:rel_acc} 

Particle acceleration at relativistic shocks, namely those shocks with Lorentz factor $\Gamma>10$, is still not clear. On general grounds, acceleration at shock front is efficient only in the case of a sizeable probability for a particle to move back to the upstream region once it crossed the shock from upstream to downstream. Only through many cycles upstream-downstream-upstream it is possible to reach efficient acceleration. 

In the case of relativistic shocks with $\Gamma\gg 1$ the shock velocity in the reference frame of the downstream plasma can be determined by imposing the conservation of energy and mass across the shock and it turns out to be $\simeq c/3$. A relativistic particle with Larmor radius $r_L$ takes a time $\tau=2\pi r_L/c$ to complete one Larmor gyration in the same time the shock moves by $\tau c/3=(2\pi/3) r_L>r_L$, therefore the particle is trapped downstream with a quite reduced probability of returning upstream (return probability). This fact leads to expect steeper spectra and reduced maximum energies for acceleration at relativistic shocks \cite{Lemoine:2006gg,Sironi:2013ri,Sironi:2015oza,Marcowith:2016vzl} in a way difficult to be reconciled with the observations of UHECR. To solve this problem, achieving the highest energies observed, one may invoke large turbulence downstream that leads to an increased return probability and harder spectra of the accelerated particles. This assumption is at the base of all models of acceleration at relativistic shocks \cite{Schneider:1987aa,Ellison:1990aa,Kirk:2000yh,Vietri:2002kq,Vietri:2003te,Blasi:2005qd}, which show a general consensus on the accelerated spectra of the type $N(E)\propto E^{-2.3}$ following from the assumption of small pitch angle scattering downstream \cite{Schneider:1987aa,Ellison:1990aa,Kirk:2000yh,Vietri:2002kq,Vietri:2003te,Blasi:2005qd}.

Assuming the required level of magnetic turbulence relativistic shocks can accelerate particles until the highest energies. The first time that a particle traverses the shock from upstream to downstream and back its energy can increase by a large factor of the order of $\simeq 4\Gamma^2$. For large values of $\Gamma$, as in GRB that can achieve $\Gamma=300$, the energy gain can be remarkable with particles that acquire energies at the level of $10^{5}\div 10^{6}$ GeV, note that this energy scale will be a low energy cutoff in the spectrum of the accelerated particles. After the first shock crossing particles distribution will be beamed within an angle of the order of $1/\Gamma$ around the shock normal. The highly anisotropic distribution of particles implies a much lower energy gain at any subsequent shock crossing, if any, of the order of $\Delta E/E\simeq 2$. Note that the simple picture discussed here applies to planar shocks, it might become somewhat different (and more complicated) for non-planar relativistic shocks as, for instance, in the case of relativistically moving plasmoids as those observed in AGN jets. 

As stated above, the key point of acceleration in relativistic shocks is related to the required magnetic field turbulence, a crucial ingredient in order to reach the energies of UHECR. Let us consider the specific case of GRB that offer an environment in which relativistic shocks are at the origin of the observed emissions. In this case the upstream magnetic field can be inferred by the observation of the X-ray afterglow emissions: typical values determined in this way are at the level of $10\div 100$ $\mu$G \cite{Li:2006ft,Li:2010zx}, significantly stronger than the interstellar value. It is tempting to attribute this apparent high magnetisation to streaming instabilities triggered by the accelerated particles penetrating the ambient plasma. However, at the present time, one cannot exclude that the circum-burst medium is per se strongly magnetised to the above level \cite{Marcowith:2016vzl,Sironi:2015oza}. Experimental evidences of the magnetic field in the downstream region, which is crucial in order to increase the return probability downstream-upstream, are more uncertain \cite{Marcowith:2016vzl,Sironi:2015oza}. Recent particle-in-cell simulations show a fast decay of the downstream magnetic turbulence that seems difficult to sustain at the required level for UHECR acceleration \cite{Sironi:2013ri,Marcowith:2016vzl,Sironi:2015oza}. 

An interesting attempt to avoid the problem of the lack of magnetic turbulence is represented by the one-shot acceleration in AGN jets proposed in \cite{Caprioli:2015zka}. In this case CR with energies around $10^{17}$ eV and below penetrate the jet sideways receiving a one-shot boost in energy of a factor $\Gamma^2$ reaching the required energies \cite{Caprioli:2015zka}.   

In the case of nuclei with large electric charge the magnetic turbulence required to attain $\simeq 10^{20}$ eV is less demanding. However, at least for GRB, the acceleration of heavy nuclei seems challenging \cite{Lemoine:2002vg,Pruet:2002hi,Horiuchi:2012by}. This is a general point connected with the expected heavy nuclei density at large red-shifts. At the cosmological epoch of reionization protons and helium nuclei were the dominant components and heavy nuclei were almost completely absent. Heavy elements provide cooling of ordinary stars during their evolution process including the pre-Super Nova phase. The stage of reionization of the universe, detected by Planck at redshift $z\lesssim10$ \cite{Adam:2016hgk}, needs at least two early generations of stars with low metallicity, Pop III and Pop II stars. Hence, the main contribution to the heavy nuclei observed in the extra-galactic space is given by the present time Super Nova explosions. This scenario is also confirmed by the observations in $Ly \alpha$ forest which indicate that the extra-galactic space had very low fraction of heavy elements at the level $Z \sim 10^{-3.5} Z_\odot$ at redshift $z \sim 5$ \cite{Songaila:2001de}. 

In conclusion, while relativistic shock acceleration has been often invoked as a viable mechanism for UHECR acceleration, especially in the context of GRB and AGN jets, there are many difficulties and poorly understood aspects, typically related with the required magnetic turbulence and mass composition, that do not allow a firm conclusion on whether this mechanism may be really at work. 

\subsection{Unipolar induction} 
\label{sec:unip_acc} 

The rotation of a magnetised star implies potentially large induced electric fields that, in turn, could accelerate particles until the highest energies of UHECR \cite{Ginzburg:1990sk}. Several different astrophysical objects show strong magnetisation, particularly suitable to accelerate ultra-high-energy particles are black hole magnetospheres and pulsars, i.e. fast spinning Neutron Stars (NS). This kind of accelerators was proposed already in 90s \cite{Venkatesan:1996jw,Blasi:2000xm,Arons:2002yj} and recently updated mainly in connection with the new observations of Auger \cite{Fang:2012rx,Fang:2013cba,Kotera:2015pya}. 

Magnetised and fast spinning NS present an energy budget which is favourable for the acceleration of UHECR, as follows from the NS population density $\dot{n}_s=3\times 10^{-3}$ Mpc$^{-3}$yr$^{-1}$ \cite{Lorimer:2008se} and the very large reservoir of rotational energy, at the level of 
\begin{equation}
E_{rot}\simeq 2\times 10^{52} \left (\frac{I}{10^{45} g cm^2}\right) \left (\frac{P}{10^{-3} s} \right)^{-2}~erg
\label{eq:Erot}
\end{equation}
with $I$ the moment of inertia and $P$ the rotation period of the star \cite{Lorimer:2008se}. Comparing these numbers with the emissivity expected from UHECR sources (see Eq. (\ref{eq:flux})), one finds that only a tiny fraction of NS, at the level of $10^{-4}$ and below, should contribute to the observed UHECR flux.  

The crust of a NS is supposed to be composed by condensed matter tightly bound in long molecular chains oriented along the magnetic field lines \cite{Ruderman:1972aj,Ruderman:1974aa}. These chains are thought to be composed by iron nuclei $^{56}Fe$ ordered in a one dimensional lattice with an outer sheath of electrons. The binding energy of iron nuclei can be estimated as $14$ keV and the lattice spacing $d\simeq 10^{-9}$ cm \cite{Ruderman:1975ju}. Therefore the electric field needed to extract an iron will be 
$${\mathcal E}_0=\frac{14 keV}{Zed}\simeq 1.4\times 10^{13} \frac{1}{Z d_{-9}}~~~\frac{V}{cm},$$ 
being $Z$ the electric charge of the nucleus extracted ($Z=26$ for iron) and $d_{-9}$ the lattice spacing in units of $10^{-9}$ cm. 

Extraction of nuclei can be achieved by the electric field generated at the NS surface by the star rotation, it can be estimated as \cite{Arons:2002yj}
\begin{equation}
{\mathcal E}=\frac{2\pi B R_s}{P c} \simeq 6.3\times 10^{14}\left (\frac{B}{10^{13} G}\right) \left (\frac{R_s}{10^{6} cm}\right) \left (\frac{P}{10^{-3} s}\right)^{-1} ~ \frac{V}{cm}
\label{eq:E_NS}
\end{equation}
where $B$ is the magnetic field at the surface of the star, $R_s$ its radius and $P$ the star rotation period. The principal effect of this field is to extract from the NS crust electrons, much less bounded than nuclei, which suffer curvature radiation in the strong dipolar magnetic field of the star's magnetosphere. Photons produced by curvature losses of electrons can in turn give rise to pairs by scattering on virtual photons associated to the magnetic field, pairs will again emit curvature photons in a cascading process. This chain of events leads to a multiplication of the number of electron-positron pairs that eventually fill the magnetosphere of the star, the number of pairs generated by a single extracted electron is in the range $10\div 10^{4}$ depending on local conditions. Pairs in the magnetosphere have a twofold effect on the acceleration of UHECR: from one side they allow milder electric fields to extract iron nuclei because of their backward acceleration toward the star, on the other side they have a screening effect on the electric field, that could be damped over large distances from the star surface. 

However, at least in principle, the total potential drop available to accelerate particles in the magnetosphere is \cite{Arons:2002yj}:
\begin{equation}
\Phi=\frac{2\pi^2 B R_s^3}{P^2 c^2}\simeq 7\times 10^{19} \left (\frac{B}{10^{13} G}\right) \left (\frac{R_s}{10^{6} cm}\right)^3 \left (\frac{P}{10^{-3} s}\right)^{-2} ~ V
\label{eq:phi_NS}
\end{equation}
which would correspond to a maximum particles' Lorentz factor $\gamma_{\Phi}=Ze\Phi/(A m_p c^2)$, i.e. an energy exceeding $10^{20}$ eV for iron nuclei. Actually, as for electrons, the maximum acceleration energy attainable will be limited by curvature losses. Assuming that the total potential drop $\Phi$ is available over a gap of length $\xi R_L$, being $R_L$ the radius of the light cylinder\footnote{Distance at which the peripheral velocity of the star reaches the velocity of light $2\pi R_L/P=c$ and the magnetic field lines spiral outwards along the azimuth.} of the star $R_L=cP/2\pi$, and equating curvature energy losses with energy gain one gets an upper bound for the acceleration Lorentz factor as \cite{Kotera:2015pya}
\begin{equation}
\gamma_{curv}=\left (\frac{3\pi B R_s^3}{2ZeP\xi c} \right)^{1/4} \simeq 10^8 \left (\frac{\xi Z}{26}\right)^{-1/4} \left (\frac{B}{10^{13} G}\right)^{1/4} \left (\frac{P}{10^{-3} s}\right)^{-1/4} \left (\frac{R_s}{10^6 cm}\right)^{3/4} ~,
\label{NS_gmax}
\end{equation}
therefore the actual maximum energy that particles can attain within the co-rotating magnetosphere will be set by $\gamma_{max}=min(\gamma_{cur},\gamma_{\Phi})$. The parameter $\xi$ takes into account the screening effect of pair creation and it can be estimated at the level of $O(1)$ \cite{Kotera:2015pya}, signalling that the gap cannot be too far from the star surface. 

If the NS wind has a Lorentz factor larger than $\gamma_{max}$ evaluated above, particles that end up in the wind will be advected with it at the Lorentz factor of the wind irrespective of the energy reached in the magnetosphere. This mechanism discussed in \cite{Kotera:2015pya} provides a way of increasing UHECR energies independently of the curvature losses.

The discussion above focuses on the most classical scenario of particles acceleration nearby the stellar surface. There are in literature other scenarios in which acceleration happens at the light cylinder or further out \cite{Cheng:1986qt} that will be not discussed here. 

The spectrum of UHECR accelerated by a NS is determined by the evolution of the rotational frequency, as the star spins down the energy of the accelerated particles decreases (see Eq. (\ref{eq:phi_NS})). Generally a NS is powered by the rotational kinetic energy and loses energy by accelerating particle winds and by emitting electromagnetic radiation. Because of energy losses the rotation frequency decreases with time, this slowdown is usually described by the relation $\dot{\nu}=-K\nu^n$ where $n$ is the braking index ($n=3$ for a pure magnetic dipole) \cite{Ruderman:1972aj} and $K$ is a positive constant, which depends on the moment of inertia and on the magnetic dipole moment of the star \cite{Ruderman:1972aj}. 

The spectrum of UHECR accelerated by the NS is given by 
$$N(E)\propto E^{\frac{1-n}{2}}~;$$ 
therefore hard as $N(E)\propto E^{-1}$ in the reference case of $n=3$ \cite{Ruderman:1972aj} and even harder for braking index $n<3$, as frequently observed.

Once nuclei are extracted from the stellar crust and accelerated by the potential gap $\Phi$ they are advected with the NS wind and interact with the environment of the star, suffering mainly photo-hadronic interactions \cite{Fang:2012rx,Fang:2013cba,Kotera:2015pya}. These are two relevant processes that can change both the energy of accelerated particles and their mass composition. 

Photo-hadronic interaction processes suffered by iron nuclei in the thermal radiation background surrounding the NS are extremely important in order to determine the actual nuclei species accelerated. As discussed in \cite{Protheroe:1997er,Kotera:2015pya} for most reasonable values of the NS surface temperature ($T<10^{7}$ $^\circ$K) a large fraction of heavy nuclei survives the losses on the radiation environment of the star. Moreover, photo-disintegration leads to the production of secondary particles (intermediate mass nuclei and nucleons, see section \ref{sec:prop}), hence, even extracting only iron nuclei from the crust of the star, the mass composition of UHECR produced by the accelerator will be anyway mixed with light nuclei and nucleons \cite{Fang:2012rx,Fang:2013cba,Kotera:2015pya}.  

\section{Exotic models} 
\label{sec:exotic}

The extreme energies of UHECR, as high as $10^{11}$ GeV, eleven orders of magnitude above the proton mass and "only" eight below the Planck mass, are a unique workbench to probe new ideas, models and theories which show their effects at energies much larger than those ever obtained, or obtainable in the future, in accelerator experiments.

There are two general classes of exotic theories that can be tested trough UHECR: top-down models for the production of these extremely energetic particles and models that imply extensions and/or violations of Lorentz invariance, as in certain theories of Quantum Gravity (QG). 

\subsection{Top-down production and super heavy dark matter}

Top-down models imply the generation of UHECR directly at the highest energies observed without passing through an acceleration mechanism that brings particles from low energy up to the highest. The idea of top-down models for the production of UHECR arose in 90s to explain the lacking of the high energy suppression in the flux observed by the AGASA experiment \cite{Takeda:1998ps}. The two main classes of top-down models are associated with topological defects and super-heavy relics (see \cite{Berezinsky:1998ft} and references therein). 

Here we do not discuss the case of topological defects that are usually associated with a symmetry breaking, as in in the case of monopoles, cosmic strings and necklaces \cite{Berezinsky:1998ft}. Nowadays, with the firm experimental observation of the high energy suppression in the UHECR flux, the scientific case for super-heavy relics is mainly connected with the problem of DM and cosmological observations. 

The leading paradigm to explain DM observations is based on the Weakly Interactive Massive Particle (WIMP) hypotheses \cite{Bergstrom:2000pn,Bertone:2004pz}, which consists of two basic assumptions: (i) WIMP are stable particles of mass $M_\chi$ (in the range $10^{2}\div 10^{4}$ GeV) that interact weakly with the Standard Model (SM) particles; (ii) WIMP are thermal relics, i.e. they were in Local Thermal Equilibrium (LTE) in the early universe. 

Searches for WIMP particles are ongoing through three different routes: direct detection, indirect detection, and accelerator searches (see \cite{Bertone:2004pz} and references therein). None of these efforts have discovered a clear WIMP candidate so far. In addition, no evidence for new physics has been observed at the Large Hadron Collider. Although not yet conclusive, the lack of evidence for WIMP may imply a different solution for the DM problem outside of the WIMP paradigm.

An alternative to WIMP models is represented by the scenario based on super-heavy relics produced due to time varying gravitational fields in the early universe: the so-called Super Heavy Dark Matter (SHDM). This approach is based on the possibility of particle production due to the non-adiabatic expansion of the background space-time acting on the vacuum quantum fluctuations. In quantum theories the possibility of producing particle pairs by the effect of a strong (classical) external field is well known: for instance, consider the case of $e^{\pm}$ pair creation by strong electromagnetic fields. The idea to apply such a mechanism in cosmology through external strong gravitational fields dates back to E. Schr\"odinger in 1939 \cite{Schro:1939}.

The construction of a coherent theory of particle creation by the expansion of the universe was developed in the last 40 years and started with the pioneering works of \cite{Chernikov:1968zm,Parker:1968mv,Grib:1970mv,Zeldovich:1971mw,Grishchuk:1974ny}. More recently, in the framework of inflationary cosmologies, it was shown that particle creation is a common phenomenon, not tied to any specific cosmological scenario, that can play a crucial role in the solution to the DM problem as SHDM (labeled by $X$) can have $\Omega_X(t_0)\simeq 1$ \cite{Chung:1998zb,Kuzmin:1998uv,Chung:1999ve,Chung:2001cb,Aloisio:2006yi,Kolb:2007vd,Fedderke:2014ura,Aloisio:2015lva}. This conclusion can be drawn under three general hypotheses: (i) SHDM in the early universe never reaches LTE; (ii) SHDM particles have mass of the order of the inflaton mass, $M_{\phi}$; and (iii) SHDM particles are long-living particles with a lifetime exceeding the age of the universe, $\tau_X\gg t_0$. 

\begin{figure}[!h]
\centering
\includegraphics[scale=0.5999]{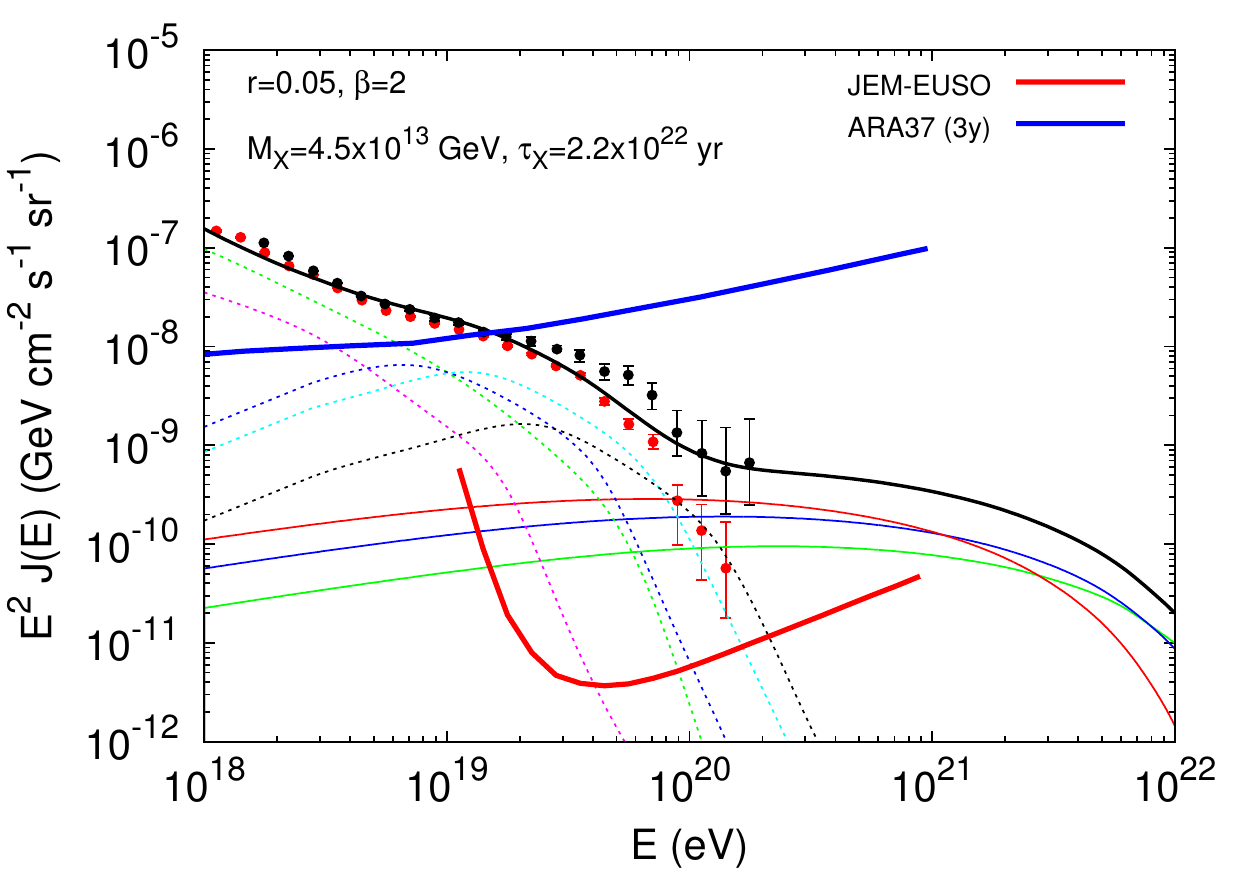}
\includegraphics[scale=0.5999]{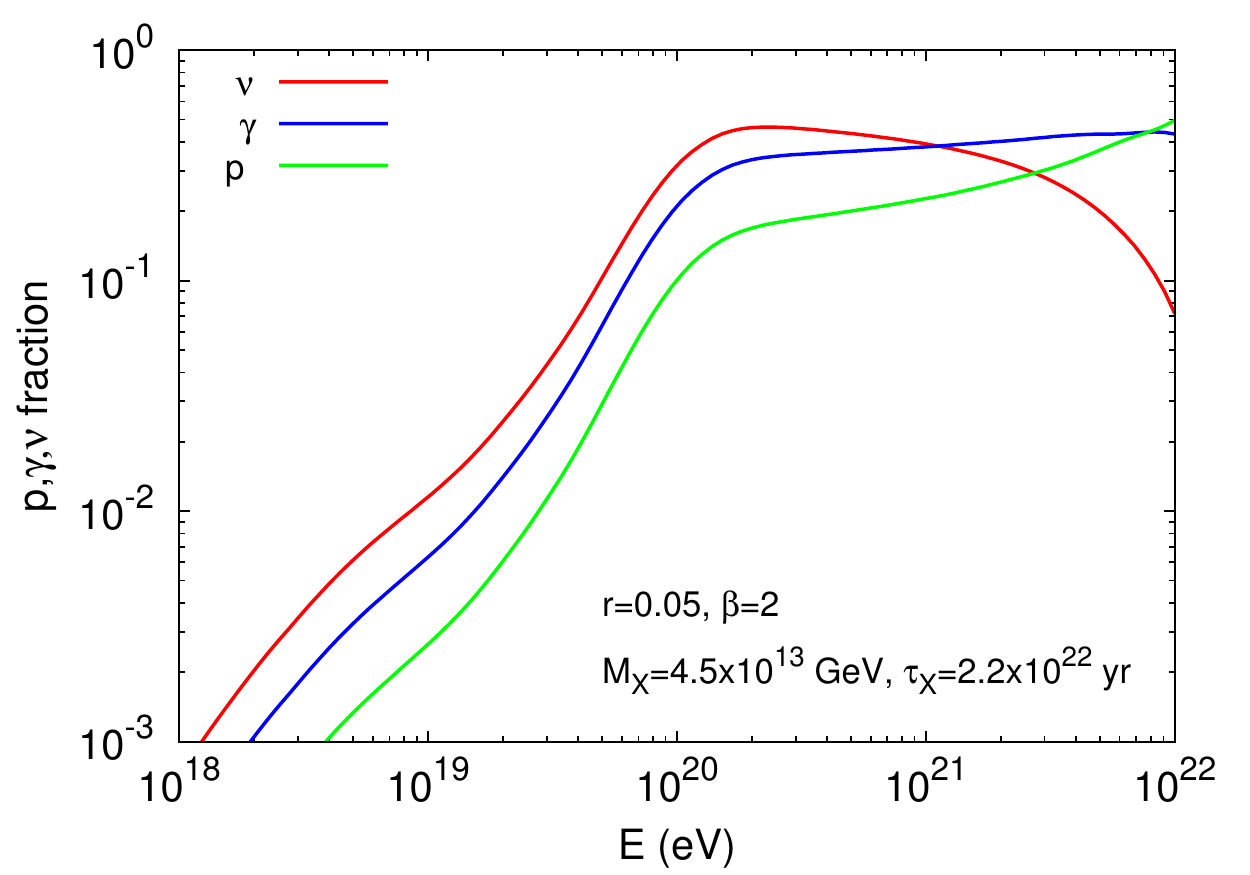}
\caption{[Left Panel] Flux of UHECR from the decay of SHDM (thin solid lines) with parameters as labeled together with the flux expected in the framework of the mixed composition model of \cite{Aloisio:2015ega}. Also shown is the sensitivity to SHDM decay products: of the proposed JEM-EUSO space mission (thick red solid line) and, for UHE neutrinos, the upcoming ARA observatory (thick blue solid line). Experimental data are those of Auger (red points) \cite{Aab:2015bza} and TA (black points) \cite{Ivanov:2015pqx}. [Right Panel] Fraction over the total flux of protons, photons and neutrinos by SHDM decay as follows from fluxes in right panel. Both figures are taken from \cite{Aloisio:2015lva}.} 
\label{fig9}   
\end{figure}

Precision measurements of CMB polarisation and observations of UHECR up to energies $\simeq 10^{20}$ eV enable a direct experimental test of the three hypothesis listed above. As discussed in \cite{Aloisio:2015lva}, the first two hypothesis can be probed through the measurements of CMB polarisation. The third hypothesis, particle life-time longer than the age of the universe, is a general requirement of any DM model based on the existence of new particles. As in the case of WIMP, discrete gauge symmetries protecting particles from fast decays work equally well for SHDM particles (see \cite{Aloisio:2003xj,Aloisio:2006yi} and references therein).

The best way to test the existence of SHDM is through the indirect detection of its annihilation and/or decay products (direct detection is unattainable). Since the annihilation cross section of a (point) particle is bounded by unitarity, $\sigma_{ann}\propto 1/M_X^2\sim 1/M_\phi^2$, the annihilation process results in an unobservable small annihilation rate \cite{Aloisio:2006yi}. Even if alternative theoretical models can be constructed, with a sizeable annihilation rate \cite{Blasi:2001hr}, we will not consider this case here and focus on the more general case of signatures from SHDM decays.

If SHDM particles decay, under general assumptions on the underlying theory (see \cite{Aloisio:2003xj,Aloisio:2006yi} and references therein), we can determine the composition and spectra of the standard model particles produced. Typical decay products are neutrinos, gamma rays and nucleons with a flat spectrum, that at the relevant energies can be approximated as $dN/dE \propto E^{-1.9}$, independently of the particle type, with a photon/nucleon ratio of about $\gamma/N\simeq 2\div 3$ and a neutrino nucleon ratio $\nu/N\simeq 3\div 4$, quite independent of the energy range \cite{Aloisio:2003xj}. The most constraining limits on SHDM lifetime are those coming from the (non) observation of UHE photons and, even to a lesser extent, neutrinos. Auger observations provide us with very stringent limits on the photons flux at energies above $10^{19}$ eV, which are at the level of $2\%$ \cite{Aglietta:2007yx}, this fact already constrains the SHDM life-time to values $\tau_X\ge 10^{21} \div 10^{22}$ yr depending on the underlying inflationary potential.

In the left panel of figure \ref{fig9}, as discussed in \cite{Aloisio:2015lva}, we plot the flux of UHECR coming from the decay of SHDM in a specific model of inflation with $M_{X}=4.5\times 10^{13}$ GeV and $\tau_X=2.2\times 10^{22}$ yr (solid lines); we also show the expected sensitivities of the proposed JEM-EUSO space mission (thick red solid line) \cite{Ebisuzaki:2014wka} and, for UHE neutrinos, of the upcoming ARA observatory (thick blue solid line) \cite{Allison:2011wk,Allison:2014kha}. In the right panel we show the corresponding fraction over the total flux of protons, photons and neutrinos by the decay of SHDM. From figure \ref{fig9} it is clear that SHDM models can be effectively probed only by the next generation of UHECR experiments, those designed to maximise statistics at the highest energies, together with new and more refined observations of the CMB polarisation pattern as they constrain the inflationary scenario.  

\subsection{Tests of Lorentz invariance} 

In the last years a general consensus has emerged on the use of UHECR observations as a powerful tool to put under experimental scrutiny Lorentz Invariance (LI). The possibility of testing LI at scales not probed so far is interesting per se, as any new experimental scrutiny of a fundamental theory, moreover the need for extensions and/or violations of LI can be connected to the long-standing problem of the construction of a quantum theory of gravity. 

Our universe is very well described by quantum mechanics at small scales and general relativity at large scales, nonetheless a unified theory of QG is still unknown. While all the other fundamental interactions propagate through space-time, gravity turns out to be a property of space-time itself. This simple statement, at the base of general relativity, has important implications for the construction of a quantum theory of gravity, as it implies that the structure of space-time itself has quantum fluctuations. In other words, at the scales where quantum effects of gravity arise, space-time is expected to have a granular (or foamy) structure where however the size of space-time cells fluctuates stochastically, thereby causing an intrinsic uncertainty in the measurements of space-time lengths, and indirectly of the energy and momentum of particles. The typical scale at which quantum gravitational effects are supposed to become relevant is the so called Planck length, i.e. the length scale given by $l_p=\sqrt{\hbar G/c^3}\simeq 1.6\times 10^{-33} $ cm. It is generally argued that measurements of distances (times) smaller than the Planck length (time) are conceptually unfeasible, since the process of measurement collects in a Planck size cell an energy exceeding the Planck mass ($M_P=\sqrt{\hbar c/G}\simeq 1.2 \times 10^{19}$ GeV) hence forming a black hole in which information is lost. 

As was immediately realised after the proposal of the GZK suppression \cite{Kirzhnits:1972sg}, in the reference frame in which astrophysical photon backgrounds are isotropic, in the case of CMB with a Planckian distribution of energies, an UHE nucleon only needs a fractional gain in energy at the level of $10^{-22}\div 10^{-21}$ to perform the transition to its final state (photo-pion production or photo-disintegration). LI guarantees that this is exactly the same to what happens in the reference frame in which the nucleon is at rest and the photon has $10\div 100$ MeV energy. But this also displays the fact that even very tiny violations of LI are bound to give, in some selected reactions at least, observable effects. The kind of reactions typically very sensitive to LI violations are those characterised by a particle production threshold \cite{Aloisio:2000cm,Aloisio:2002ed,Aloisio:2005rc,Aloisio:2006nd}. 

In recent times LI violating models have been deeply analysed and compared with available experimental data \cite{Liberati:2013xla,Kostelecky:2008ts}. Particularly interesting is the approach of Effective Field Theories (EFT) in which LI or CPT symmetry (and renormalizability) is no longer a guide, in this kind of theories the number of possible terms violating LI is very high. Those that can be tested experimentally (several hundreds) are describe in \cite{Kostelecky:2008ts} and can be generally modelled through modifications of the dispersion relation of particles \cite{Carmona:2012un} (which in the EFT approach corresponds to modifications of kinetic terms in the Lagrangian density), such as: 
\begin{equation}
E^2-p^2=\mu^2(E,p,M_P)\simeq m^2 + f E^{2+n}/M_P^n
\label{eq:mashell}
\end{equation}
where $\mu$ is a generalised "mass" that can be always approximated as the mass of the particle $m$ plus terms that violate LI at the strength fixed by $f$ ($f=0$ corresponds to the standard invariant relation). 

The firm experimental evidence of the suppression in the spectrum of UHECR around a few$\times 10^{19}$ eV implies very stringent limits on the possible violations of LI. Using the parameterisation introduced in Eq. (\ref{eq:mashell}), the case $f>0$ is strongly excluded by observations, because in this case thresholds for particles' production move to lower energies and new exotic processes are allowed, such as vacuum Cerenkov $p\to p\gamma$ for which very strong bounds exists \cite{Klinkhamer:2008ss}. As soon as $f$ moves toward negative values, thresholds for particles' production slightly increase until the point at which become kinematically forbidden. In this case, limits of LI violations obtained from the observed spectral steepening are reported in literature \cite{Jacobson:2005bg,Saveliev:2011vw}. These limits, however, depend crucially on the assumption that the steepening in the flux is originated by the propagation of UHECR. As discussed in section \ref{sec:prop}, Auger data can be very well accommodated in models in which the flux suppression is connected with low values (see section \ref{sec:mix}) of the maximum acceleration energy at the source. In this case no relevant limit on LI violations can be deduced using the observed flux of UHECR \cite{Aloisio:2014dua}. 

Violations of LI can also imply important effects in the development of showers produced by the interaction of UHECR with nuclei of the Earth's atmosphere. These effects typically reduce the kinematical phase space for certain processes modifying the particles content of the cascading shower. The most important process in the physics of cascades is the neutral pion decay $\pi^0\to \gamma\gamma$, which has a reduced kinematic phase space in the case of LI violations ($f<0$) with stable neutral pions at energies larger than $E>(M_P^n m_\pi^2/|f|)^{\frac{1}{2+n}}$ \cite{Aloisio:2014dua}. This modification of particles' cascade has the net effect to move the shower maximum to higher altitudes as the electromagnetic part of the shower consumes faster. Mover, it produces an increased number of high energy muons in the shower due to the interaction of "non-decaying" neutral pions. As of today, observations of the showers development in the atmosphere are not able to exclude LI violations effects, that are however much weaker and difficult to tag than in the case of UHECR propagation. 

\section{Discussion and conclusions}
\label{sec:conclu}

The most important physical task in the physics of UHECR is certainly connected with a clear identification of the sources. As always in cosmic ray physics, the observables that we can use to solve the mystery are: flux, mass composition and, if any, anisotropy of the arrival directions. In the present paper we concentrated the discussion on the first two observables, showing how the flux can be interpreted in different ways with a certain level of degeneracy due to mass composition and that the latter is still unclear with a high degree of uncertainty connected with the actual measurements of Auger and TA.

As discussed in sections \ref{sec:intro}, \ref{sec:prop} and \ref{sec:models}, mass composition has a paramount importance in the physics of propagation, it fixes few fundamental characteristics of the sources ($\gamma_g$ and $E_{max}$) and it regulates the amount of secondary (observable) particles, gamma rays and neutrinos, that, in a multi-messenger approach, could unveil important informations about sources. 

The experimental determination of mass composition in both Auger and TA is based on the observation of fluorescence light produced by the electromagnetic component of the EAS in the atmosphere. These measurements, performed through the fluorescence telescopes, enable the determination of the atmospheric depth at which the number of particles in the EAS reaches its maximum $X_{max}(E)$, which is directly related to the number of nucleons of the primary particle hitting the atmosphere. Heavy nuclei interact higher in the atmosphere with smaller fluctuations. In practice, the actual quantity which can be measured and relates to mass composition is the distribution $N(X_{max})$ of the $X_{max}$ position of showers with total energy $E$. 

In the past, because of the limited statistic, the observation of $N(X_{max})$ was restricted to the first two moments of the distribution, $\langle X_{\max} \rangle$ and $\sigma(X_{\max})$ respectively, as discussed in section \ref{sec:models}. This kind of analysis performed by both Auger and TA brings to the uncertainties discussed and seems not conclusive \cite{Aab:2014kda,Abbasi:2015xga}. 
 
Recently, the Auger collaboration, exploiting its high statistics, published an analysis based on the reconstruction of the whole behaviour of the distribution $N(X_{max})$ at different energies \cite{Aab:2016zth,Aab:2014aea}. The comparison of observations with Monte Carlo simulations provides a more precise determination of the mass composition. The results of \cite{Aab:2014aea} are quite interesting, even if not conclusive and not incompatible with previous analysis, seem to point toward a lesser heavy composition respect to the one extracted from the first two moments of the distribution. However, also in this case, a large level of uncertainty comes from the hadronic interaction model implemented in the Monte Carlo simulations \cite{Aab:2014aea}. 

The measurement of mass composition only by means of fluorescence seems not sufficient, another observable which plays an important role in tagging composition is the number of muons observed by the surface detectors. The basic effect to distinguish a nucleus from a proton by observing muons is related to the different energy per nucleon, $E/A$, at fixed total energy $E$. A low energy nucleon produces low energy charged pions which decay to muons before the parent pion undergoes new collisions with air-nuclei. 

The Auger experiment has several unique possibilities to measure the muon flux directly and use it to determine the mass composition. The on-ground water-Cherenkov detectors (tanks) can measure muons in inclined directions, with some level of uncertainty due to the decoupling of the electron and muon signals. Moreover, the AugerPrime \cite{Aab:2016vlz} upgrade has  been  specifically  designed  to  improve muons measurements in the whole energy range. Along the line of a hybrid design, each tank will be equipped with scintillator layer on the top. Shower particles will be sampled by two detectors (scintillators and water-Cherenkov) having different responses to the muonic and electromagnetic components, thus allowing to reconstruct each of them separately. The muonic component will be derived in each station by subtracting the signal observed in the scintillator from that seen in the water Cherenkov tank. The upgraded array will provide data with no duty cycle limitation allowing the access to the highest energies.

As discussed in \cite{Younk:2012mp}, the precise measurement of the muon production depth $X_{\max}^{\mu}$, i.e. the atmospheric depth corresponding to the maximum production of muons, may give valuable informations if correlated with $X_{\max}$, measured by fluorescent light. A preliminary analysis of this correlation, performed on the Auger muon data extracted from inclined showers, points toward a mixed composition at the highest energies $\gtrsim 10^{19}$ eV with, at least, three nuclei components among which the iron group seems to be excluded \cite{Aab:2016htd}. 

Given the strong joint efforts of Auger and TA collaborations in solving the puzzle of mass composition, it is reasonable to expect that in the near future a clearer view will emerge. Meanwhile, in the present review, we have considered the two alternative scenarios of light or heavy composition to deliver a wider view of the physics of UHECR. In section \ref{sec:models} we have considered the dip model, which assumes a pure proton composition and implies steep injection spectra $\gamma_g\gtrsim2.5$ and high maximum energies of acceleration $E_{max}\gtrsim 10^{20}$ eV, and the mixed composition model, which shows a richer phenomenology with heavy elements injected with hard spectra $\gamma\lesssim1.5$ and light elements with soft spectra $\gamma_g\gtrsim2.5$ both with relatively low maximum acceleration energies $E_{max}\lesssim Z\times 5\times 10^{18}$ eV. 

As discussed in section \ref{sec:source}, a pure proton composition is, theoretically, a natural possibility. Proton is the most abundant element in the universe and several different astrophysical objects, at present and past cosmological epochs, could provide efficient acceleration even if it requires very high luminosities and maximum acceleration energies. The latter being particularly challenging for many models of magnetic turbulence in both cases of relativistic and non-relativistic shocks.

The complexity of the scenario based on a composition with heavy nuclei disfavours astrophysical sources placed at high redshift because of the lacking of heavy elements. As discussed in section \ref{sec:source}, mixed composition favours sources with flat injection spectra for heavy nuclei, provided by acceleration mechanisms based on strong electric fields, such those at the surface of rotating neutron stars (see section \ref{sec:unip_acc}), or invoking particular dynamics in the source environment, such as photo-disintegration on strong local photon fields \cite{Globus:2014fka,Unger:2015laa}. A remarkable feature of the mixed composition scenario is the relative low maximum energy required at the source.

The study of secondary cosmogenic particles produced by the propagation of UHECR is another fundamental aspect of this field of research. As discussed in section \ref{sec:sec}, UHE neutrinos and gamma/electron/positron are always produced by photo-hadron interactions involving UHECR. While the latter are rapidly absorbed through ICS and PP processes, contributing to the diffuse gamma ray background ($\lesssim 100$ GeV), the former bring direct informations about sources and mass composition up to the highest red-shifts. 

Theoretical expectations on the fluxes of secondary messengers are very sensitive to the mass composition. In the case of pure protons there is a huge production of both neutrinos and gamma rays. As discussed in section \ref{sec:sec}, particularly constraining is the diffuse gamma ray background as recently detected by the Fermi-LAT satellite \cite{TheFermi-LAT:2015ykq}, that seems to exclude certain source models and their cosmological evolution in the framework of a pure proton composition \cite{Liu:2016brs,Berezinsky:2016jys}. 

After the first detection of astrophysical neutrinos at energies $\lesssim 10^{15}$ eV by the IceCube collaboration \cite{Aartsen:2013jdh,Aartsen:2013bka}, the study of HE and UHE neutrinos attracted a renewed interest. The observations of IceCube, being at relatively low energy, can be only marginally explained in the framework of cosmogenic neutrinos coming from UHECR interactions, also given the large uncertainties on the EBL background at high red-shift \cite{Aloisio:2015ega}. At high energies ($E\gtrsim 10^{18}$) neutrino production is critically related with the mass composition of UHECR and with cosmological evolution of sources. Therefore the lacking of neutrino observations in this energy regime is not too constraining for UHECR models. 

As discussed in section \ref{sec:exotic}, the study of UHECR has an impact also in fundamental physics, as it involves tests of models and theories that extend beyond the standard model of particle physics. This is the case of SHDM that, being a viable alternative to the WIMP paradigm for DM, can be tested only through the combined observations of the CMB polarisation pattern and UHECR at the highest energies \cite{Aloisio:2015lva}. 

We conclude highlighting the two principal avenues on which the study of UHECR should develop in the near future. From one hand, as discussed above, a firm experimental determination of the mass composition is an unavoidable step forward in this field of research. On the other hand, the highest energy regime, typically the trans-GZK energies $E \gtrsim 5\times 10^{19}$ eV, still remains less probed with not enough statistics to firmly detect possible anisotropies in the arrival directions and the exact shape of the suppression. Current technologies can reach one order of magnitude more in the number of observed events at the highest energies, which seems not enough to firmly detect anisotropies or to probe new physics as the SHDM hypothesis. New technologies are needed and future space observatories, with improved photon detection techniques, like JEM-EUSO \cite{Ebisuzaki:2014wka}, OWL \cite{Cline:1999ez} or Super-EUSO \cite{Petrolini:2009cg}, promise a new era in the physics of UHECR. 

\section*{Acknowledgments}
The author is grateful to P. Blasi, V. Berezinsky, D. Boncioli, I. De Mitri, A. Di Matteo, C. Evoli, A. Gazizov, S. Grigorieva, S. Matarrese, G. Morlino, A. Olinto, S. Petrera and F. Salamida for useful discussions and continuous collaboration on these and related research topics. The author wants to dedicate the present work to the memory of Aurelio F. Grillo, not only an outstanding colleague, who introduced many young scientist to the physics of cosmic rays, but also and first of all a close friend.


\end{document}